\documentclass[superscriptaddress,amsmath,amssymb,aps,prl,twocolumn,showpacs,nofootinbib]{revtex4-1}
\pdfoutput=1
\usepackage{float}
\usepackage{graphicx}
\usepackage{dcolumn}
\usepackage{url}

\begin{document}

\title{Manipulation of nonequilibrium spin dynamics of an ultracold gas in a moving optical lattice}

\author{Z.~N. Hardesty-Shaw}
\affiliation{Department of Physics, Oklahoma State University, Stillwater, Oklahoma 74078, USA}

\author{Q. Guan}
\affiliation{Homer L. Dodge Department of Physics and Astronomy, The University of Oklahoma, Norman,
Oklahoma 73019, USA}
\affiliation{Center for Quantum Research and Technology, The University of Oklahoma, Norman, Oklahoma 73019, USA}
\affiliation{Department of Physics and Astronomy, Washington State University, Pullman, WA 99164, USA}

\author{J.~O. Austin}
\affiliation{Department of Physics, Oklahoma State University, Stillwater, Oklahoma 74078, USA}

\author{D. Blume}
\affiliation{Homer L. Dodge Department of Physics and Astronomy, The University of Oklahoma, Norman,
Oklahoma 73019, USA}
\affiliation{Center for Quantum Research and Technology, The University of Oklahoma, Norman, Oklahoma 73019, USA}

\author{R.~J. Lewis-Swan}
\email{lewisswan@ou.edu} \affiliation{Homer L. Dodge Department of Physics and Astronomy, The University of Oklahoma,
Norman, Oklahoma 73019, USA} \affiliation{Center for Quantum Research and Technology, The University of Oklahoma, Norman,
Oklahoma 73019, USA}

\author{Y. Liu}
\email{yingmei.liu@okstate.edu} \affiliation{Department of Physics, Oklahoma State University, Stillwater, Oklahoma 74078,
USA}

\date{\today}

\begin{abstract}
The isolation and control of disparate degrees of freedom underpin quantum simulators. We advance the programmability of cold atom quantum simulators with a first realization of the dynamic interplay of spatial and spin degrees of freedom. We experimentally demonstrate that violent spatial evolutions tune long-lived coherent spin dynamics and develop a model of quantum spin-mixing incorporating the spatial evolution via time-dependent spin-spin interactions. Our results open new paths towards the simulation of quantum spin models with tunable interactions via tailored spatial dynamics.
\end{abstract}

\maketitle

\noindent{\it Introduction --}
Ultracold quantum gases that feature spatial and spin degrees of freedom offer a powerful platform for simulating quantum magnetism in controlled, isolated settings~\cite{polkovnikov2011colloquium,bloch2008many,stamper2013spinor,lewenstein2007ultracold,Eckert2007spinorlattice}.
When combined with optical lattices, these simulation capabilities are exemplified by experimental studies featuring tunable dimensionality and filling factors \cite{zhao2015antiferromagnetic,austin2021pra,austin2021quantum,chen2019quantum}.
Possessing long coherence times, these systems also provide an ideal platform for studying out-of-equilibrium phenomena such as spin-mixing \cite{Becker2010spinorlattice,zhao2014dynamics,Mahmud2013spinmixing,gabardos2020dipolar}, transport \cite{gross2014spintransport,Jepsen2020transport}, dynamical phases of matter \cite{thywissen2019dpt}, and critical dynamics across quantum phase transitions \cite{Jiang2016mott,austin2021pra,austin2021quantum}. Simultaneously, advances in spin- and spatially-resolved probes \cite{boll2016spindensity,Asteria2021magnifier,chen2019quantum,austin2021pra} and the control of time- and spin-dependent lattice potentials
are opening up new opportunities, including the study of multi-state tunneling physics \cite{Sinitsyn_2017} and driven-dissipative phases \cite{esslinger2019spinor}, in the presence of the spin degree of freedom.

Typically, the energy scales of the spin and spatial degrees of freedom are disparate. This has been exploited to obtain a reduced description of the spin dynamics that depends only on a spatial profile that remains frozen due to, e.g., a strong confining potential~\cite{law1998spinmixing,pu1999spin,yi2002single,amr2013spinmodel,amr2014sun,krauser2014giant,thywissen2019dpt,zhao2014dynamics,liu2009quantum}.
In the context of spinor Bose-Einstein condensates (BECs), this decoupled regime has received significant attention \cite{sengstock2006spinresonance,pechkis2013thermal,jie2020mean} and, amongst other applications, has been utilized to generate entangled states in the highly controllable spin degree of freedom \cite{gross2011atomic,klempt2011twinatoms,hamley2012spin,peise2015satisfying,mengkhoon2018dicke,qu2020squeezing}, which can also be mapped to the motional degrees of freedom \cite{klempt2021momentum,qingze2021mz,oberthaler2018spatialEPR}.

In contrast, the interplay of spatial and spin degrees of freedom remains largely unexplored, and, although typically weak, can provide a powerful avenue for controlling the spin dynamics through tailored dynamical manipulation of the spatial properties of the gas
\cite{klempt2018spatialEPR,kunkel2019noncommuting,kunkel2022entagstructure,santos2014spinchain,zinner2015spinchain}.
We investigate this interplay, providing a first example of how spatial degrees of freedom can be utilized to manipulate the spin dynamics. We experimentally observe that a one-dimensional (1D) moving lattice, combined with a skew optical dipole trap (ODT), induces violent transient spatial motion which is nevertheless accompanied by long-lived spin-mixing dynamics.
We develop a theoretical understanding of these observations based on a \emph{dynamical single spatial-mode approximation} (dSMA), which leads to an effective spin model with a time-dependent spin-spin interaction coefficient that depends on the temporal evolution of the BEC density profile.
Experimental observations --- including a robust critical regime featuring divergent timescales for the spin dynamics, which is tuned by the applied moving optical lattice and associated spatial motion --- are qualitatively described by our model.

Our results open the way for the exploitation of classical spatial dynamics for simulating many-body quantum spin dynamics with highly tunable, time-dependent interactions \cite{davis2019spin1,periwal2021programmable}, thereby enhancing the class of quantum spin models accessible in spinor BECs. In addition, our findings imply spatial dynamics can provide new control knobs for the nonequilibrium generation of entangled spin states for, e.g., quantum-enhanced sensing ~\cite{klempt2021momentum}.

\begin{figure*}[ht!]
\includegraphics[width=\textwidth]{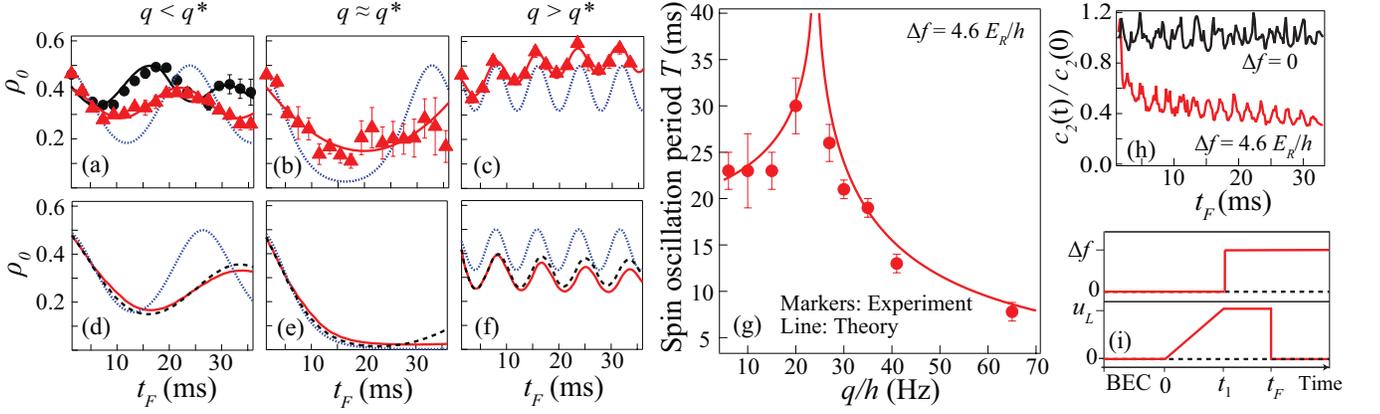}
\centering \caption{\label{fig:fig1}
(a)-(f) Exemplary time traces of $\rho_0$ for spinor BECs. Panels (a)-(c) show experimental results for $\rho_0$ (red markers) at $q/h = 15$~Hz (a), $25$~Hz (b), and $65$~Hz (c) as well as sSMA predictions (dotted lines) with $c_{2,\mathrm{fit}}/h = 23.7(1)$Hz extracted from Fig.~\ref{fig:fig1}(g). Panel (a) compares static (black) and moving (red) lattice results. Solid lines are sinusoidal fits to guide the eye.
Panels (d)-(f) compare predicted $\rho_0$ for $q/h = 15$~Hz (d), $22$~Hz (e), and $65$~Hz (f) from $1$D GP simulations (solid lines) to sSMA (dotted lines) with $c_2$ obtained through fits to the GP data~\cite{sm2022}, and dSMA (dashed lines) with $c_2(t)$ obtained from GP data. 
The chosen $q$ values exemplify the interaction dominated ($q < q^*$) and Zeeman ($q > q^*$) regimes separated by the critical region $q \approx q^*$. 
(g) Observed $T$ (markers) versus $q$ fit by analytical sSMA expressions (solid line) with the fitting parameter $c_{2,\mathrm{fit}}/h = 23.7(1)$Hz~\cite{sm2022}.
(h) Red (black) lines are evolution of $c_2(t)/c_2(0)$ for $q/h = 15$~Hz obtained from $1$D GP simulations of the  moving (static) lattice.
All moving lattice data use a lattice depth $u_L = 2.3E_R$, $t_1 = 1.43$ms, and fixed $\Delta f = 4.6E_R/h$.
(i)  Timeline of the lattice depth (lower panel) and moving lattice speed (upper panel).}
\end{figure*}

\begin{figure}[t]
\includegraphics[width=\linewidth]{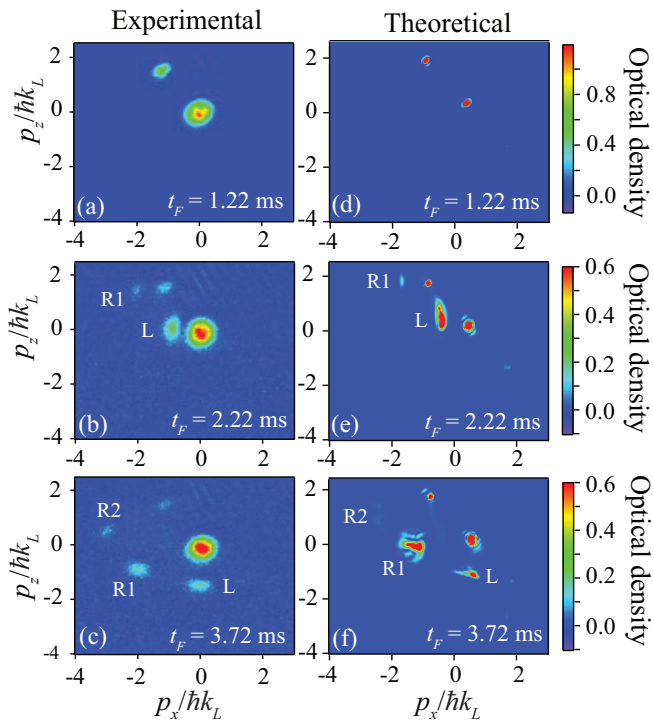}
\centering \caption{\label{fig:fig2}
Time-of-flight snapshots of $2$D integrated momentum distribution with $\Delta f= 4.6E_R/h$, $u_L = 1.2 E_R$, $t_1 = 0.72$~ms, and $t_F = 1.22$~ms (a), $2.22$~ms (b), and $3.72$~ms (c) in a shallow ODT with $L_{\mathrm{ODT,z}}=20~\mu\mathrm{m}$~\cite{sm2022} at $q/h = 42$Hz. (d)-(f) Analogous theoretical results of in-situ momentum distributions based on $2$D GP simulations~\cite{sm2022}. The colorbar scale indicates the optical density of the images for each row.}
\end{figure}

\noindent{\it Experimental setup --}
Each experimental cycle begins with a sodium spin-1 BEC
at quadratic Zeeman energy $q$ in an ODT
(see Supplemental Materials~\cite{sm2022}). A key
feature of spinor BECs is their spin degree of freedom characterized by the spin-dependent interaction coefficient $c_2$.
Spin-mixing and other nonequilibrium phenomena driven by a static $c_2$ have been studied in various
contexts~\cite{jie2020mean,pu1999spin,yi2002single,austin2021quantum, austin2021pra,zhao2014dynamics,liu2009quantum}. Here, in contrast, we demonstrate that $c_2$ can be tuned dynamically by utilizing a moving lattice to change the BEC's spatial density profile. We construct a 1D moving lattice with two nearly orthogonal optical beams whose
frequency difference $\Delta f$ determines the moving lattice speed~\cite{sm2022}. Our initial BEC has a
fractional population $\rho_0 = 0.5$ of atoms in the $|S=1, m=0\rangle$ state and zero magnetization (equal populations
in the $|S=1,m = \pm 1\rangle$ states). The BEC is then adiabatically loaded into the lattice, which is stationary at time $t=0$ and quenched to the desired speed for $t>0$ (see Fig.~\ref{fig:fig1}(i)). We study the ensuing non-trivial spin (Fig.~\ref{fig:fig1}) and spatial (Fig.~\ref{fig:fig2}) dynamics of the atoms by holding them in the moving lattice for a time $t_F$ before releasing them for ballistic expansion and imaging~\cite{chen2019quantum,sm2022}.

\noindent{\it Spin dynamics --}
We first study the nonequilibrium spin dynamics generated by experimental sequences (Fig.~\ref{fig:fig1}(i) $\Delta f = 4.6 E_R/h$) which near-resonantly couple the initial stationary BEC with the
$\mathbf{p}=2\hbar\mathbf{k}_L$ momentum state.
Here, $E_R$ is the recoil energy, $h$ ($\hbar$) is the (reduced) Planck
constant, and $\mathbf{k}_L$ is the lattice vector~\cite{zhao2015antiferromagnetic,sm2022}. 
Spin-mixing oscillations, arising from coherent interconversions among two $m=0$ atoms and a pair of atoms in the $m=\pm1$ Zeeman
states~\cite{stamper2013spinor}, constitute a useful tool in understanding the spin dynamics.
The periods $T$ of these oscillations are determined by the competition between $c_2$ and $q$, illustrated by typical
examples of the interaction dominated region (Fig.~\ref{fig:fig1}(a)) and Zeeman dominated region
(Fig.~\ref{fig:fig1}(c)). We also see convincing experimental signatures of a critical separatrix regime near $q = q^*$ where $T$ diverges (see Fig.~\ref{fig:fig1}(b)). These observations are qualitatively
consistent with expectations based on established theory formulations referred to as a static single spatial-mode
approximation (sSMA) in this paper, which assumes $c_2$ is time-independent
\cite{stamper2013spinor,jie2020mean,pu1999spin,yi2002single,austin2021quantum,
austin2021pra,zhao2014dynamics,liu2009quantum}. Fig.~\ref{fig:fig1}(g) shows the observed
$T$ can be used to estimate the effective static spin-spin interaction $c_{2,\mathrm{fit}}/h = 23.7(1)$~Hz as sSMA predicts $q^* \approx c_{2,\mathrm{fit}}$ for our initial
state~\cite{zhao2015antiferromagnetic,zhao2014dynamics}. 
However, direct comparisons to the sSMA predicted time traces (dotted lines in Figs.~\ref{fig:fig1}(a)-(c)) demonstrate that the model fails to capture experimentally observed
features such as the damping of the oscillation amplitude and the drift of the oscillations. 
Another notable observation that cannot be explained by sSMA is the shift of the separatrix location induced by the moving lattice, as shown by a comparison between static ($\Delta f = 0$) and moving lattice results in Fig.~\ref{fig:fig1}(a).\protect{\footnote{Fig.~\ref{fig:fig1}(a) indicates that for static lattices at an identical $q$ in the interaction dominated regime $T$ is smaller and thus, using the same sSMA interpretation, would lie on a curve shifted to higher $q$ (indicating a larger characteristic $c_2$) relative to the moving lattice data shown in Fig.~\ref{fig:fig1}(g).}}
These experimental observations suggest sSMA provides an incomplete description of our system.

\noindent{\it Theoretical model --}
To explain the sSMA's shortcomings, we develop the following dSMA model which assumes c2 varies with time and describes our system with the spin Hamiltonian~\cite{law1998spinmixing,qingze2021mz,zhao2015antiferromagnetic,sm2022},
\begin{equation}\label{eqn:Ham}
    \hat{H}_{\rm eff}(t) = \frac{c_2(t)}{2N}\hat{\mathbf{S}} \cdot \hat{\mathbf{S}} + q(\hat{n}_1 + \hat{n}_{-1} ) .
\end{equation}
Here, $\hat{\mathbf{S}} = \sum_{i=1}^N \hat{\mathbf{s}}_i$ where $\hat{\mathbf{s}}_i$ denotes the spin-$1$ operator for the $i$th of the total N atoms and $\hat{n}_m$ is the number operator for the Zeeman state $m$.
The time-dependent $c_2(t)$ arises from the temporal evolution of the BEC's spatial density profile and, in turn, modulation of the effective interaction strength of the spin model, driven by the moving lattice. Formally, $c_2(t)$ emerges from the time-dependence of the Gross-Pitaevskii (GP) orbitals $\psi_m({\mathbf{r}},t)$ that describe the spatial dynamics of the $m$th Zeeman component. By assuming the spatial density profiles $|\psi_m({\mathbf{r}},t)|^2$ for the different $m$ states are the same but time-dependent (as we find in theory calculations discussed in more detail later)
it is implicit that while the spatial degree of freedom may contribute to the spin dynamics the converse is not true, i.e., the spin degree of freedom does not feed back onto the evolution of the spatial profile. 
Setting $|\psi_m({\mathbf{r}},t)|^2 \propto \vert\phi({\mathbf{r}},t)|^2$ and integrating out the spatial degrees of freedom leads to Eq.~(\ref{eqn:Ham}) with
$c_2(t) \propto (N-1) \int d^{3}\mathbf{r}~\vert \phi(\mathbf{r},t) \vert^4$~\cite{sm2022}.
A key result of the presented experiment-theory work is that dSMA enables a transparent understanding of the non-trivial spin dynamics triggered by violent spatial evolution of the BEC that occurs on faster characteristic time scales than the spin dynamics.

\noindent{\it Interplay of spin and spatial dynamics --} 
To illustrate the typical spatial dynamics driving the spin-mixing observed in Fig.~\ref{fig:fig1}, we show experimental BEC momentum distributions in Figs.~\ref{fig:fig2}(a)-(c), which  capture the emergence of violent spatial motion due to the interplay of momentum kicks generated by the moving lattice and the shallow ODT harmonic confinement on a timescale significantly shorter than the observed spin dynamics.

The rapid appearance of many of discrete momentum peaks and associated spatial dynamics shown in
Fig.~\ref{fig:fig2} simultaneously suggests that the deviations from sSMA predictions in
Figs.~\ref{fig:fig1}(a)-(c) are to be expected but also entices us to reconcile elements of the good qualitative agreement
between the experimental data and sSMA calculations in Fig.~\ref{fig:fig1}(g). 
We note that the creation of the discrete momentum peaks is a coherent process and does not conflict with the assumption of the single spatial-mode approximation. In fact, the Supplemental Materials~\cite{sm2022} show that Figs.~\ref{fig:fig1}(a)-(c) are replicated if different momentum components are used to construct $\rho_0$, thereby providing experimental support for dSMA. 

To gain further insight, we use numerical GP calculations~\cite{sm2022}, which provides a mean-field description of the full spinor BEC dynamics including both spatial and spin degrees of freedom. The complexity of the experimental system, in particular the disparate
timescales of spin and spatial dynamics, precludes a full quantitative $3$D GP treatment. Instead, we use a reduced
dimensionality $1$D spinor GP calculation with parameters tuned to capture essential aspects of the experimental $3$D
system. This simplified treatment enables us to develop a qualitative understanding of the experimental results~\cite{sm2022}. The GP simulations (solid lines in Figs.~\ref{fig:fig1}(d)-(f))
qualitatively replicate the coherent spin dynamics, including a diverging oscillation period for
$q \approx q^*$ (Fig.~\ref{fig:fig1}(e)) and robust harmonic oscillations for $q < q^{*}$ (Fig.~\ref{fig:fig1}(d)) and $q
> q^{*}$ (Fig.~\ref{fig:fig1}(f)) with damped amplitude and drifting mean value, respectively. We note that the upward
drifting mean value in the experimental data for $q > q^{*}$ (Fig.~\ref{fig:fig1}(c)), notably not captured by the 1D GP
theory, may be induced by a subtle resonance mechanism between the spin and spatial dynamics~\cite{jie2020mean} that
depends sensitively on the dimensionality of the system. Higher dimensionality calculations will be presented elsewhere~\cite{futurePaper}.

We use GP calculations to make a more fine-grained theoretical investigation of the relationship between the spatial and spin dynamics and, in particular, certify that while the BEC undergoes violent motion on fast time-scales: i) all Zeeman components are described by a common spatial density profile $\vert \phi(\mathbf{r},t)\vert^2$, and ii) the sophisticated interplay of the moving lattice and ODT drive complex dynamics of $c_2(t)$. These observations lead us to self-consistently compare the $1$D GP results to dSMA predictions, i.e., mean-field dynamics based on Eq.~(\ref{eqn:Ham}) with $c_2(t)$ computed via the GP density $|\phi(\mathbf{r},t)|^2$~\cite{sm2022}. The GP and dSMA time traces in Figs.~\ref{fig:fig1}(d)-(f) show excellent agreement with each other. This implicitly demonstrates the spin dynamics do not feed back into the spatial evolution, in agreement with expectations based on the disparity of energy scales. The dSMA results also show significantly improved qualitative agreements with experimental data than sSMA results, providing further support for our dSMA model.

GP calculations of the spatial dynamics in the presence of a moving lattice lead to appreciable variation of $c_2(t)$ (red line in Fig.~\ref{fig:fig1}(h)) at $t \lesssim 3$~ms. Over longer timescales $c_2(t)$ features an overall decrease, which we understand as being driven by the relaxation of the spatial density profile as the BEC fractures into many momentum components.
This behavior is in stark contrast with predictions for a static lattice (black line in Fig.~\ref{fig:fig1}(h)) that indicate $c_2(t)$ instead fluctuates around a well defined time-averaged value with small oscillations due to excitations created during the loading phase.
After the initial transient behavior in the moving lattice, our results (Fig.~\ref{fig:fig1}(h)) indicate that the decay of $c_2(t)$ is slow relative to the characteristic time of the spin dynamics, and therefore observables such as the spin oscillation period are captured by sSMA~\cite{sm2022}.
Our calculations show that the precise details of the spin dynamics (e.g., qualitative features including damping of the spin oscillations in Fig.~\ref{fig:fig1}(a)) can depend greatly on the temporal variation of $c_2(t)$ and hence a more rigorous description is provided by the GP and dSMA models.

\begin{figure}[t]
\includegraphics[width=\linewidth]{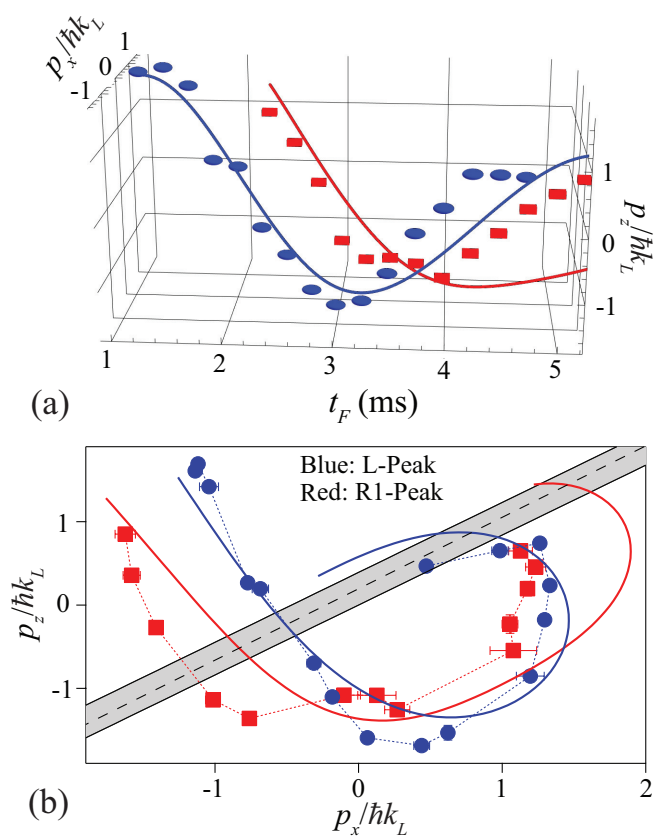}
\centering \caption{\label{fig:fig3} 
(a) Time evolution trajectories of the mean position of the L-peak (circles) and R1-peak (squares) in the $p_z$-$p_x$ plane taken in a compressed ODT with $L_{\mathrm{ODT,z}}=33~\mu\mathrm{m}$~\cite{sm2022}, extracted from experimental images similar to those shown in Figs.~\ref{fig:fig2}(a)-(c) at $q/h = 42$Hz. With increasing time, markers appear larger due to movement in the $p_x$ axis.
Blue (red) solid lines are the L-peak (R1-peak) trajectories corresponding to Lissajous curves (see text)~\cite{sm2022}. (b) Trajectories of panel (a) projected into the $p_z$-$p_x$ plane. The gray shaded region encompasses an approximate resonance region where decelerated atoms are kicked by the lattice (to create, e.g., the R1-peak)~\cite{sm2022}.
}
\end{figure}

\noindent{\it Single-particle resonances --}
The precise evolution of the spatial density profile, seen in
Figs.~\ref{fig:fig2}(a)-(c) as a menagerie of seemingly irregularly distributed wavepackets in momentum space, can be
understood with the aid of GP simulations (Figs.~\ref{fig:fig2}(d)-(f)). To more precisely capture the impact of
gravity and the finite trap depth, the simulations in Fig.~\ref{fig:fig2} are performed using an axially symmetric $2$D setup. This numerical treatment is feasible due to the relatively
short time scales over which the spatial dynamics are studied in detail. Using $2$D simulations also enables us to capture
key details of the momentum kicks that $1$D simulations, such as those employed in Fig.~\ref{fig:fig1}, miss. At short
times, the lattice kicks atoms from the initial BEC with momentum $\mathbf{p} = 0$ to the near-resonant state with
momentum $\mathbf{p}=2\hbar \mathbf{k}_L$  (Figs.~\ref{fig:fig2}(a) and \ref{fig:fig2}(d)). Subsequently, an additional
momentum component, referred to as the lead peak (L-peak), splits from the $\mathbf{p}=2\hbar \mathbf{k}_L$ peak and
decelerates as it travels away from the minima of the relatively shallow ODT potential. As the L-peak slows, its momentum
evolves until it sweeps through the approximate resonance region centered on the line $p_z \approx 0.86p_x+0.20\hbar k_L$
(see the gray shaded region in Fig.~\ref{fig:fig3}(b) and Supplemental Materials~\cite{sm2022}), where the lattice couples
two nearly resonant momentum states, corresponding to the L-peak and a new peak labelled R1, that are separated by another
$2\hbar\mathbf{k}_L$ momentum kick (Figs.~\ref{fig:fig2}(b) and \ref{fig:fig2}(e)). Similarly, the R1-peak also
decelerates until it crosses the resonance region and the lattice generates a new peak labelled R2
(Figs.~\ref{fig:fig2}(c) and \ref{fig:fig2}(f)). This pattern continues and the BEC fractures into a multitude of momentum
states.

Figure \ref{fig:fig3}(a) confirms our prior analysis, which suggests a dependence on both the confining ODT potential and moving lattice, by tracking the position of the L- and R1-peaks in time and momentum space in a compressed ODT.
The position of the L-peak initially follows a trajectory consistent with a Lissajous curve derived from a simplified classical treatment of a single particle initially moving with momentum $2\hbar\mathbf{k}_L$ in the ODT (see Supplemental Materials~\cite{sm2022}). A sudden momentum kick imparted by the lattice couples the L- and R1-peaks as the former crosses the approximate resonance region, shown by the gray region in Fig.~\ref{fig:fig3}(b)~\cite{sm2022}.
As time increases (i.e., deeper into the trajectory of each peak), the agreement between the experimentally observed and the theoretically predicted trajectories for the L- and R1-peaks deteriorates, as shown near the end of the time axis in Fig.~\ref{fig:fig3}(a).
The deterioration is most clearly seen when comparing an ideal (effectively $L_{\rm ODT,i} = \infty$) harmonic trap with our typical shallow ODT depth that features a comparatively small curvature (see Supplemental Materials~\cite{sm2022}).
In Fig.~\ref{fig:fig3}, we use compressed ODTs instead of shallow ODTs, trading reduced visibility and condensate fraction, to better illustrate the bending of the trajectory in the $p_x-p_z$ plane~\cite{sm2022}.

\noindent{\it Conclusion --}
Our results open a new direction for the exploitation of spatial dynamics as a control knob for quantum simulations of many-body quantum spin models with tunable, time-dependent interactions \cite{davis2019spin1,periwal2021programmable}. Tailored modulation of the spatial profile could be used to control the precise time dependence of the spin-spin interactions and realize Floquet-driven spin dynamics \cite{fujimoto2019floquetspinor,PhysRevA.100.033617}. This can have immediate applications for the dynamical generation of entangled spin states for quantum-enhanced sensing
\cite{mengkhoon2018dicke,feldmann2018qpt,mirkhalaf2020sensing,qingze2021mz,rls2022optical}. The observed short time dynamics also raise intriguing questions about equilibration of spinor BECs. For example, future studies might utilize the time dependence of $c_2(t)$ to force these systems along different equilibration trajectories. 

\begin{acknowledgments}
\noindent{\it Acknowledgements --} D.~B. acknowledges support by the National Science Foundation (NSF) through grant No.
PHY-2110158. R.~J. L-S. acknowledges support by NSF through Grant No. PHY-2110052 and the Dodge Family College of Arts and
Sciences at the University of Oklahoma (OU). Z.~N.~H-S., J.~O.~A., and Y.~L. acknowledge support by the Noble Foundation
and the NSF through Grant Nos. PHY-1912575 and PHY-2207777. This work used  the OU Supercomputing Center for Education and
Research (OSCER).
\end{acknowledgments}


\begin{thebibliography}{100}%
\makeatletter
\providecommand \@ifxundefined [1]{%
 \@ifx{#1\undefined}
}%
\providecommand \@ifnum [1]{%
 \ifnum #1\expandafter \@firstoftwo
 \else \expandafter \@secondoftwo
 \fi
}%
\providecommand \@ifx [1]{%
 \ifx #1\expandafter \@firstoftwo
 \else \expandafter \@secondoftwo
 \fi
}%
\providecommand \natexlab [1]{#1}%
\providecommand \enquote  [1]{``#1''}%
\providecommand \bibnamefont  [1]{#1}%
\providecommand \bibfnamefont [1]{#1}%
\providecommand \citenamefont [1]{#1}%
\providecommand \href@noop [0]{\@secondoftwo}%
\providecommand \href [0]{\begingroup \@sanitize@url \@href}%
\providecommand \@href[1]{\@@startlink{#1}\@@href}%
\providecommand \@@href[1]{\endgroup#1\@@endlink}%
\providecommand \@sanitize@url [0]{\catcode `\\12\catcode `\$12\catcode
  `\&12\catcode `\#12\catcode `\^12\catcode `\_12\catcode `\%12\relax}%
\providecommand \@@startlink[1]{}%
\providecommand \@@endlink[0]{}%
\providecommand \url  [0]{\begingroup\@sanitize@url \@url }%
\providecommand \@url [1]{\endgroup\@href {#1}{\urlprefix }}%
\providecommand \urlprefix  [0]{URL }%
\providecommand \Eprint [0]{\href }%
\providecommand \doibase [0]{http://dx.doi.org/}%
\providecommand \selectlanguage [0]{\@gobble}%
\providecommand \bibinfo  [0]{\@secondoftwo}%
\providecommand \bibfield  [0]{\@secondoftwo}%
\providecommand \translation [1]{[#1]}%
\providecommand \BibitemOpen [0]{}%
\providecommand \bibitemStop [0]{}%
\providecommand \bibitemNoStop [0]{.\EOS\space}%
\providecommand \EOS [0]{\spacefactor3000\relax}%
\providecommand \BibitemShut  [1]{\csname bibitem#1\endcsname}%
\let\auto@bib@innerbib\@empty


\bibitem{polkovnikov2011colloquium}
\bibinfo{author}{Polkovnikov, A.}, \bibinfo{author}{Sengupta, K.},
  \bibinfo{author}{Silva, A.} \& \bibinfo{author}{Vengalattore, M.}
\newblock \bibinfo{title}{Colloquium: Nonequilibrium dynamics of closed
  interacting quantum systems}.
\newblock \emph{\bibinfo{journal}{Rev. Mod. Phys.}}
  \textbf{\bibinfo{volume}{83}}, \bibinfo{pages}{863} (\bibinfo{year}{2011}).

\bibitem{bloch2008many}
\bibinfo{author}{Bloch, I.}, \bibinfo{author}{Dalibard, J.} \&
  \bibinfo{author}{Zwerger, W.}
\newblock \bibinfo{title}{Many-body physics with ultracold gases}.
\newblock \emph{\bibinfo{journal}{Rev. Mod. Phys.}}
  \textbf{\bibinfo{volume}{80}}, \bibinfo{pages}{885} (\bibinfo{year}{2008}).

\bibitem{stamper2013spinor}
\bibinfo{author}{Stamper-Kurn, D.~M.} \& \bibinfo{author}{Ueda, M.}
\newblock \bibinfo{title}{Spinor {B}ose gases: Symmetries, magnetism, and
  quantum dynamics}.
\newblock \emph{\bibinfo{journal}{Rev. Mod. Phys.}}
  \textbf{\bibinfo{volume}{85}}, \bibinfo{pages}{1191} (\bibinfo{year}{2013}).

\bibitem{lewenstein2007ultracold}
\bibinfo{author}{Lewenstein, M.} \emph{et~al.}
\newblock \bibinfo{title}{Ultracold atomic gases in optical lattices: mimicking
  condensed matter physics and beyond}.
\newblock \emph{\bibinfo{journal}{Adv. Phys.}} \textbf{\bibinfo{volume}{56}},
  \bibinfo{pages}{243--379} (\bibinfo{year}{2007}).

\bibitem{Eckert2007spinorlattice}
\bibinfo{author}{Eckert, K.}, \bibinfo{author}{Zawitkowski, {\L}.},
  \bibinfo{author}{Leskinen, M.~J.}, \bibinfo{author}{Sanpera, A.} \&
  \bibinfo{author}{Lewenstein, M.}
\newblock \bibinfo{title}{Ultracold atomic {B}ose and {F}ermi spinor gases in
  optical lattices}.
\newblock \emph{\bibinfo{journal}{New J. Phys.}} \textbf{\bibinfo{volume}{9}},
  \bibinfo{pages}{133--133} (\bibinfo{year}{2007}).
\newblock \urlprefix\url{https://doi.org/10.1088/1367-2630/9/5/133}.

\bibitem{zhao2015antiferromagnetic}
\bibinfo{author}{Zhao, L.}, \bibinfo{author}{Jiang, J.}, \bibinfo{author}{Tang,
  T.}, \bibinfo{author}{Webb, M.} \& \bibinfo{author}{Liu, Y.}
\newblock \bibinfo{title}{Antiferromagnetic spinor condensates in a
  two-dimensional optical lattice}.
\newblock \emph{\bibinfo{journal}{Phys. Rev. Lett.}}
  \textbf{\bibinfo{volume}{114}}, \bibinfo{pages}{225302}
  (\bibinfo{year}{2015}).

\bibitem{austin2021pra}
\bibinfo{author}{Austin, J.~O.}, \bibinfo{author}{Shaw, Z.~N.},
  \bibinfo{author}{Chen, Z.}, \bibinfo{author}{Mahmud, K.~W.} \&
  \bibinfo{author}{Liu, Y.}
\newblock \bibinfo{title}{Manipulating atom-number distributions and detecting
  spatial distributions in lattice-confined spinor gases}.
\newblock \emph{\bibinfo{journal}{Phys. Rev. A}}
  \textbf{\bibinfo{volume}{104}}, \bibinfo{pages}{L041304}
  (\bibinfo{year}{2021}).
\newblock
  \urlprefix\url{https://link.aps.org/doi/10.1103/PhysRevA.104.L041304}.

\bibitem{austin2021quantum}
\bibinfo{author}{Austin, J.~O.}, \bibinfo{author}{Chen, Z.},
  \bibinfo{author}{Shaw, Z.~N.}, \bibinfo{author}{Mahmud, K.~W.} \&
  \bibinfo{author}{Liu, Y.}
\newblock \bibinfo{title}{Quantum critical dynamics in a spinor {Hubbard} model
  quantum simulator}.
\newblock \emph{\bibinfo{journal}{Commun. Phys.}} \textbf{\bibinfo{volume}{4}},
  \bibinfo{pages}{61} (\bibinfo{year}{2021}).

\bibitem{chen2019quantum}
\bibinfo{author}{Z. Chen, T. Tang, J. Austin, Z. Shaw, L. Zhao, Y. Liu},
\newblock \bibinfo{title}{Quantum quench and nonequilibrium dynamics in
  lattice-confined spinor condensates}.
\newblock \emph{\bibinfo{journal}{Phys. Rev. Lett.}}
  \textbf{\bibinfo{volume}{123}}, \bibinfo{pages}{113002}
  (\bibinfo{year}{2019}).

\bibitem{Becker2010spinorlattice}
\bibinfo{author}{Becker, C.} \emph{et~al.}
\newblock \bibinfo{title}{Ultracold quantum gases in triangular optical
  lattices}.
\newblock \emph{\bibinfo{journal}{New J. Phys.}} \textbf{\bibinfo{volume}{12}},
  \bibinfo{pages}{065025} (\bibinfo{year}{2010}).
\newblock \urlprefix\url{https://doi.org/10.1088/1367-2630/12/6/065025}.

\bibitem{zhao2014dynamics}
\bibinfo{author}{Zhao, L.}, \bibinfo{author}{Jiang, J.}, \bibinfo{author}{Tang,
  T.}, \bibinfo{author}{Webb, M.} \& \bibinfo{author}{Liu, Y.}
\newblock \bibinfo{title}{Dynamics in spinor condensates tuned by a microwave
  dressing field}.
\newblock \emph{\bibinfo{journal}{Phys. Rev. A}} \textbf{\bibinfo{volume}{89}},
  \bibinfo{pages}{023608} (\bibinfo{year}{2014}).

\bibitem{Mahmud2013spinmixing}
\bibinfo{author}{Mahmud, K.~W.} \& \bibinfo{author}{Tiesinga, E.}
\newblock \bibinfo{title}{Dynamics of spin-1 bosons in an optical lattice: Spin
  mixing, quantum-phase-revival spectroscopy, and effective three-body
  interactions}.
\newblock \emph{\bibinfo{journal}{Phys. Rev. A}} \textbf{\bibinfo{volume}{88}},
  \bibinfo{pages}{023602} (\bibinfo{year}{2013}).
\newblock \urlprefix\url{https://link.aps.org/doi/10.1103/PhysRevA.88.023602}.

\bibitem{gabardos2020dipolar}
\bibinfo{author}{L. Gabardos, B. Zhu, S. Lepoutre, A. M. Rey, B. Laburthe-Tolra, and L. Vernac},
\newblock \bibinfo{title}{Relaxation of the collective magnetization of a dense
  3d array of interacting dipolar $s=3$ atoms}.
\newblock \emph{\bibinfo{journal}{Phys. Rev. Lett.}}
  \textbf{\bibinfo{volume}{125}}, \bibinfo{pages}{143401}
  (\bibinfo{year}{2020}).
\newblock
  \urlprefix\url{https://link.aps.org/doi/10.1103/PhysRevLett.125.143401}.

\bibitem{gross2014spintransport}
\bibinfo{author}{S. Hild, T. Fukuhara, P. Schauss,J. Zeiher, M. Knap, E. Demler, I. Bloch, and C.Gross},
\newblock \bibinfo{title}{Far-from-equilibrium spin transport in {H}eisenberg
  quantum magnets}.
\newblock \emph{\bibinfo{journal}{Phys. Rev. Lett.}}
  \textbf{\bibinfo{volume}{113}}, \bibinfo{pages}{147205}
  (\bibinfo{year}{2014}).
\newblock
  \urlprefix\url{https://link.aps.org/doi/10.1103/PhysRevLett.113.147205}.

\bibitem{Jepsen2020transport}
\bibinfo{author}{Jepsen, P.~N.} \emph{et~al.}
\newblock \bibinfo{title}{Spin transport in a tunable {H}eisenberg model
  realized with ultracold atoms}.
\newblock \emph{\bibinfo{journal}{Nature}} \textbf{\bibinfo{volume}{588}},
  \bibinfo{pages}{403--407} (\bibinfo{year}{2020}).
\newblock \urlprefix\url{https://doi.org/10.1038/s41586-020-3033-y}.

\bibitem{thywissen2019dpt}
\bibinfo{author}{Smale, S.} \emph{et~al.}
\newblock \bibinfo{title}{Observation of a transition between dynamical phases
  in a quantum degenerate {F}ermi gas}.
\newblock \emph{\bibinfo{journal}{Sci. Adv.}} \textbf{\bibinfo{volume}{5}},
  \bibinfo{pages}{eaax1568} (\bibinfo{year}{2019}).
\newblock
  \urlprefix\url{https://www.science.org/doi/abs/10.1126/sciadv.aax1568}.
\newblock \eprint{https://www.science.org/doi/pdf/10.1126/sciadv.aax1568}.

\bibitem{Jiang2016mott}
\bibinfo{author}{J. Jiang, L. Zhao, S.-T. Wang, Z. Chen, T. Tang, L.-M. Duan, and Y. Liu},
\newblock \bibinfo{title}{First-order superfluid-to-mott-insulator phase
  transitions in spinor condensates}.
\newblock \emph{\bibinfo{journal}{Phys. Rev. A}} \textbf{\bibinfo{volume}{93}},
  \bibinfo{pages}{063607} (\bibinfo{year}{2016}).
\newblock \urlprefix\url{https://link.aps.org/doi/10.1103/PhysRevA.93.063607}.

\bibitem{boll2016spindensity}
\bibinfo{author}{Boll, M.} \emph{et~al.}
\newblock \bibinfo{title}{Spin- and density-resolved microscopy of
  antiferromagnetic correlations in {F}ermi-{H}ubbard chains}.
\newblock \emph{\bibinfo{journal}{Science}} \textbf{\bibinfo{volume}{353}},
  \bibinfo{pages}{1257--1260} (\bibinfo{year}{2016}).
\newblock
  \urlprefix\url{https://www.science.org/doi/abs/10.1126/science.aag1635}.
\newblock \eprint{https://www.science.org/doi/pdf/10.1126/science.aag1635}.

\bibitem{Asteria2021magnifier}
\bibinfo{author}{Asteria, L.}, \bibinfo{author}{Zahn, H.~P.},
  \bibinfo{author}{Kosch, M.~N.}, \bibinfo{author}{Sengstock, K.} \&
  \bibinfo{author}{Weitenberg, C.}
\newblock \bibinfo{title}{Quantum gas magnifier for sub-lattice-resolved
  imaging of 3d quantum systems}.
\newblock \emph{\bibinfo{journal}{Nature}} \textbf{\bibinfo{volume}{599}},
  \bibinfo{pages}{571--575} (\bibinfo{year}{2021}).
\newblock \urlprefix\url{https://doi.org/10.1038/s41586-021-04011-2}.

\bibitem{Sinitsyn_2017}
\bibinfo{author}{Sinitsyn, N.~A.} \& \bibinfo{author}{Chernyak, V.~Y.}
\newblock \bibinfo{title}{The quest for solvable multistate {L}andau-{Z}ener
  models}.
\newblock \emph{\bibinfo{journal}{J. Phys. A Math. Theor.}}
  \textbf{\bibinfo{volume}{50}}, \bibinfo{pages}{255203}
  (\bibinfo{year}{2017}).
\newblock \urlprefix\url{https://doi.org/10.1088/1751-8121/aa6800}.

\bibitem{esslinger2019spinor}
\bibinfo{author}{Dogra, N.} \emph{et~al.}
\newblock \bibinfo{title}{Dissipation-induced structural instability and chiral
  dynamics in a quantum gas}.
\newblock \emph{\bibinfo{journal}{Science}} \textbf{\bibinfo{volume}{366}},
  \bibinfo{pages}{1496--1499} (\bibinfo{year}{2019}).
\newblock
  \urlprefix\url{https://www.science.org/doi/abs/10.1126/science.aaw4465}.
\newblock \eprint{https://www.science.org/doi/pdf/10.1126/science.aaw4465}.

\bibitem{law1998spinmixing}
\bibinfo{author}{Law, C.~K.}, \bibinfo{author}{Pu, H.} \&
  \bibinfo{author}{Bigelow, N.~P.}
\newblock \bibinfo{title}{{Quantum spins mixing in spinor {B}ose-{E}instein
  condensates}}.
\newblock \emph{\bibinfo{journal}{Phys. Rev. Lett.}}
  \textbf{\bibinfo{volume}{81}}, \bibinfo{pages}{5257--5261}
  (\bibinfo{year}{1998}).
\newblock \urlprefix\url{https://link.aps.org/doi/10.1103/PhysRevLett.81.5257}.

\bibitem{pu1999spin}
\bibinfo{author}{Pu, H.}, \bibinfo{author}{Law, C.~K.},
  \bibinfo{author}{Raghavan, S.}, \bibinfo{author}{Eberly, J.~H.} \&
  \bibinfo{author}{Bigelow, N.~P.}
\newblock \bibinfo{title}{Spin-mixing dynamics of a spinor {B}ose-{E}instein
  condensate}.
\newblock \emph{\bibinfo{journal}{Phys. Rev. A}} \textbf{\bibinfo{volume}{60}},
  \bibinfo{pages}{1463--1470} (\bibinfo{year}{1999}).
\newblock \urlprefix\url{https://link.aps.org/doi/10.1103/PhysRevA.60.1463}.

\bibitem{yi2002single}
\bibinfo{author}{Yi, S.}, \bibinfo{author}{M\"ustecapl\ifmmode \imath \else \i
  \fi{}o\ifmmode~\breve{g}\else \u{g}\fi{}lu, O.~E.}, \bibinfo{author}{Sun,
  C.~P.} \& \bibinfo{author}{You, L.}
\newblock \bibinfo{title}{Single-mode approximation in a spinor-1 atomic
  condensate}.
\newblock \emph{\bibinfo{journal}{Phys. Rev. A}} \textbf{\bibinfo{volume}{66}},
  \bibinfo{pages}{011601(R)} (\bibinfo{year}{2002}).
\newblock \urlprefix\url{https://link.aps.org/doi/10.1103/PhysRevA.66.011601}.

\bibitem{amr2013spinmodel}
\bibinfo{author}{Martin, M.~J.} \emph{et~al.}
\newblock \bibinfo{title}{A quantum many-body spin system in an optical lattice
  clock}.
\newblock \emph{\bibinfo{journal}{Science}} \textbf{\bibinfo{volume}{341}},
  \bibinfo{pages}{632--636} (\bibinfo{year}{2013}).
\newblock
  \urlprefix\url{https://www.science.org/doi/abs/10.1126/science.1236929}.
\newblock \eprint{https://www.science.org/doi/pdf/10.1126/science.1236929}.

\bibitem{amr2014sun}
\bibinfo{author}{Zhang, X.} \emph{et~al.}
\newblock \bibinfo{title}{Spectroscopic observation of su(\textit{N})-symmetric
  interactions in {Sr} orbital magnetism}.
\newblock \emph{\bibinfo{journal}{Science}} \textbf{\bibinfo{volume}{345}},
  \bibinfo{pages}{1467--1473} (\bibinfo{year}{2014}).
\newblock
  \urlprefix\url{https://www.science.org/doi/abs/10.1126/science.1254978}.
\newblock \eprint{https://www.science.org/doi/pdf/10.1126/science.1254978}.

\bibitem{krauser2014giant}
\bibinfo{author}{Krauser, J.~S.} \emph{et~al.}
\newblock \bibinfo{title}{Giant spin oscillations in an ultracold {F}ermi sea}.
\newblock \emph{\bibinfo{journal}{Science}} \textbf{\bibinfo{volume}{343}},
  \bibinfo{pages}{157--160} (\bibinfo{year}{2014}).

\bibitem{liu2009quantum}
\bibinfo{author}{Y. Liu, S. Jung, S. E. Maxwell, L. D. Turner, E. Tiesinga, and P. D. Lett},
\newblock \bibinfo{title}{Quantum phase transitions and continuous observation
  of spinor dynamics in an antiferromagnetic condensate}.
\newblock \emph{\bibinfo{journal}{Phys. Rev. Lett.}}
  \textbf{\bibinfo{volume}{102}}, \bibinfo{pages}{125301}
  (\bibinfo{year}{2009}).

\bibitem{sengstock2006spinresonance}
\bibinfo{author}{Kronj\"ager, J.}, \bibinfo{author}{Becker, C.},
  \bibinfo{author}{Navez, P.}, \bibinfo{author}{Bongs, K.} \&
  \bibinfo{author}{Sengstock, K.}
\newblock \bibinfo{title}{Magnetically tuned spin dynamics resonance}.
\newblock \emph{\bibinfo{journal}{Phys. Rev. Lett.}}
  \textbf{\bibinfo{volume}{97}}, \bibinfo{pages}{110404}
  (\bibinfo{year}{2006}).
\newblock
  \urlprefix\url{https://link.aps.org/doi/10.1103/PhysRevLett.97.110404}.

\bibitem{pechkis2013thermal}
\bibinfo{author}{H. K. Pechkis, J. P. Wrubel, A. Schwettmann, P. F. Griffin, R. Barnett, E. Tiesinga, and P. D. Lett},
\newblock \bibinfo{title}{Spinor dynamics in an antiferromagnetic spin-1
  thermal {B}ose gas}.
\newblock \emph{\bibinfo{journal}{Phys. Rev. Lett.}}
  \textbf{\bibinfo{volume}{111}}, \bibinfo{pages}{025301}
  (\bibinfo{year}{2013}).
\newblock
  \urlprefix\url{https://link.aps.org/doi/10.1103/PhysRevLett.111.025301}.

\bibitem{jie2020mean}
\bibinfo{author}{Jie, J.}, \bibinfo{author}{Guan, Q.}, \bibinfo{author}{Zhong,
  S.}, \bibinfo{author}{Schwettmann, A.} \& \bibinfo{author}{Blume, D.}
\newblock \bibinfo{title}{{Mean-field spin-oscillation dynamics beyond the
  single-mode approximation for a harmonically trapped spin-1 {B}ose-{E}instein
  condensate}}.
\newblock \emph{\bibinfo{journal}{Phys. Rev. A}}
  \textbf{\bibinfo{volume}{102}}, \bibinfo{pages}{023324}
  (\bibinfo{year}{2020}).
\newblock \urlprefix\url{https://link.aps.org/doi/10.1103/PhysRevA.102.023324}.

\bibitem{gross2011atomic}
\bibinfo{author}{Gross, C.} \emph{et~al.}
\newblock \bibinfo{title}{Atomic homodyne detection of continuous-variable
  entangled twin-atom states}.
\newblock \emph{\bibinfo{journal}{Nature}} \textbf{\bibinfo{volume}{480}},
  \bibinfo{pages}{219--223} (\bibinfo{year}{2011}).

\bibitem{klempt2011twinatoms}
\bibinfo{author}{Lücke, B.} \emph{et~al.}
\newblock \bibinfo{title}{Twin matter waves for interferometry beyond the
  classical limit}.
\newblock \emph{\bibinfo{journal}{Science}} \textbf{\bibinfo{volume}{334}},
  \bibinfo{pages}{773--776} (\bibinfo{year}{2011}).
\newblock
  \urlprefix\url{https://www.science.org/doi/abs/10.1126/science.1208798}.
\newblock \eprint{https://www.science.org/doi/pdf/10.1126/science.1208798}.

\bibitem{hamley2012spin}
\bibinfo{author}{Hamley, C.~D.}, \bibinfo{author}{Gerving, C.},
  \bibinfo{author}{Hoang, T.}, \bibinfo{author}{Bookjans, E.} \&
  \bibinfo{author}{Chapman, M.~S.}
\newblock \bibinfo{title}{Spin-nematic squeezed vacuum in a quantum gas}.
\newblock \emph{\bibinfo{journal}{Nat. Phys.}} \textbf{\bibinfo{volume}{8}},
  \bibinfo{pages}{305--308} (\bibinfo{year}{2012}).

\bibitem{peise2015satisfying}
\bibinfo{author}{Peise, J.} \emph{et~al.}
\newblock \bibinfo{title}{Satisfying the {E}instein--{P}odolsky--{R}osen
  criterion with massive particles}.
\newblock \emph{\bibinfo{journal}{Nat. Commun.}} \textbf{\bibinfo{volume}{6}},
  \bibinfo{pages}{1--8} (\bibinfo{year}{2015}).

\bibitem{mengkhoon2018dicke}
\bibinfo{author}{Zou, Y.-Q.} \emph{et~al.}
\newblock \bibinfo{title}{Beating the classical precision limit with spin-1
  {D}icke states of more than 10,000 atoms}.
\newblock \emph{\bibinfo{journal}{Proc. Natl. Acad. Sci. U.S.A.}}
  \textbf{\bibinfo{volume}{115}}, \bibinfo{pages}{6381--6385}
  (\bibinfo{year}{2018}).

\bibitem{qu2020squeezing}
\bibinfo{author}{Qu, A.}, \bibinfo{author}{Evrard, B.},
  \bibinfo{author}{Dalibard, J.} \& \bibinfo{author}{Gerbier, F.}
\newblock \bibinfo{title}{Probing spin correlations in a {B}ose-{E}instein
  condensate near the single-atom level}.
\newblock \emph{\bibinfo{journal}{Phys. Rev. Lett.}}
  \textbf{\bibinfo{volume}{125}}, \bibinfo{pages}{033401}
  (\bibinfo{year}{2020}).
\newblock
  \urlprefix\url{https://link.aps.org/doi/10.1103/PhysRevLett.125.033401}.

\bibitem{klempt2021momentum}
\bibinfo{author}{F. Anders, A. Idel, P. Feldmann, D. Bondarenko, S. Loriani, K. Lange,} \emph{et~al.}
\newblock \bibinfo{title}{Momentum entanglement for atom interferometry}.
\newblock \emph{\bibinfo{journal}{Phys. Rev. Lett.}}
  \textbf{\bibinfo{volume}{127}}, \bibinfo{pages}{140402}
  (\bibinfo{year}{2021}).
\newblock
  \urlprefix\url{https://link.aps.org/doi/10.1103/PhysRevLett.127.140402}.

\bibitem{qingze2021mz}
\bibinfo{author}{Guan, Q.}, \bibinfo{author}{Biedermann, G.~W.},
  \bibinfo{author}{Schwettmann, A.} \& \bibinfo{author}{Lewis-Swan, R.~J.}
\newblock \bibinfo{title}{Tailored generation of quantum states in an entangled
  spinor interferometer to overcome detection noise}.
\newblock \emph{\bibinfo{journal}{Phys. Rev. A}}
  \textbf{\bibinfo{volume}{104}}, \bibinfo{pages}{042415}
  (\bibinfo{year}{2021}).
\newblock \urlprefix\url{https://link.aps.org/doi/10.1103/PhysRevA.104.042415}.

\bibitem{oberthaler2018spatialEPR}
\bibinfo{author}{Kunkel, P.} \emph{et~al.}
\newblock \bibinfo{title}{Spatially distributed multipartite entanglement
  enables epr steering of atomic clouds}.
\newblock \emph{\bibinfo{journal}{Science}} \textbf{\bibinfo{volume}{360}},
  \bibinfo{pages}{413--416} (\bibinfo{year}{2018}).
\newblock
  \urlprefix\url{https://www.science.org/doi/abs/10.1126/science.aao2254}.
\newblock \eprint{https://www.science.org/doi/pdf/10.1126/science.aao2254}.

\bibitem{klempt2018spatialEPR}
\bibinfo{author}{Lange, K.} \emph{et~al.}
\newblock \bibinfo{title}{Entanglement between two spatially separated atomic
  modes}.
\newblock \emph{\bibinfo{journal}{Science}} \textbf{\bibinfo{volume}{360}},
  \bibinfo{pages}{416--418} (\bibinfo{year}{2018}).
\newblock
  \urlprefix\url{https://www.science.org/doi/abs/10.1126/science.aao2035}.
\newblock \eprint{https://www.science.org/doi/pdf/10.1126/science.aao2035}.

\bibitem{kunkel2019noncommuting}
\bibinfo{author}{P. Kunkel, M. Prufer, S. Lannig, R. Rosa-Medina, A. Bonnin, M. Garttner, H. Strobel, and M. K. Oberthaler},
\newblock \bibinfo{title}{Simultaneous readout of noncommuting collective spin
  observables beyond the standard quantum limit}.
\newblock \emph{\bibinfo{journal}{Phys. Rev. Lett.}}
  \textbf{\bibinfo{volume}{123}}, \bibinfo{pages}{063603}
  (\bibinfo{year}{2019}).
\newblock
  \urlprefix\url{https://link.aps.org/doi/10.1103/PhysRevLett.123.063603}.

\bibitem{kunkel2022entagstructure}
\bibinfo{author}{P. Kunkel, M. Prufer, S. Lannig, R. Strohmaier, M. Garttner, H. Strobel, and M. K. Oberthaler},
\newblock \bibinfo{title}{Detecting entanglement structure in continuous
  many-body quantum systems}.
\newblock \emph{\bibinfo{journal}{Phys. Rev. Lett.}}
  \textbf{\bibinfo{volume}{128}}, \bibinfo{pages}{020402}
  (\bibinfo{year}{2022}).
\newblock
  \urlprefix\url{https://link.aps.org/doi/10.1103/PhysRevLett.128.020402}.

\bibitem{santos2014spinchain}
\bibinfo{author}{Deuretzbacher, F.}, \bibinfo{author}{Becker, D.},
  \bibinfo{author}{Bjerlin, J.}, \bibinfo{author}{Reimann, S.~M.} \&
  \bibinfo{author}{Santos, L.}
\newblock \bibinfo{title}{Quantum magnetism without lattices in strongly
  interacting one-dimensional spinor gases}.
\newblock \emph{\bibinfo{journal}{Phys. Rev. A}} \textbf{\bibinfo{volume}{90}},
  \bibinfo{pages}{013611} (\bibinfo{year}{2014}).
\newblock \urlprefix\url{https://link.aps.org/doi/10.1103/PhysRevA.90.013611}.

\bibitem{zinner2015spinchain}
\bibinfo{author}{A. G. Volosniev, D. Petrosyan, M. Valiente, D. V. Fedorov, A. S. Jensen, and N. T. Zinner},
\newblock \bibinfo{title}{Engineering the dynamics of effective spin-chain
  models for strongly interacting atomic gases}.
\newblock \emph{\bibinfo{journal}{Phys. Rev. A}} \textbf{\bibinfo{volume}{91}},
  \bibinfo{pages}{023620} (\bibinfo{year}{2015}).
\newblock \urlprefix\url{https://link.aps.org/doi/10.1103/PhysRevA.91.023620}.

\bibitem{davis2019spin1}
\bibinfo{author}{Davis, E.~J.}, \bibinfo{author}{Bentsen, G.},
  \bibinfo{author}{Homeier, L.}, \bibinfo{author}{Li, T.} \&
  \bibinfo{author}{Schleier-Smith, M.~H.}
\newblock \bibinfo{title}{Photon-mediated spin-exchange dynamics of spin-1
  atoms}.
\newblock \emph{\bibinfo{journal}{Phys. Rev. Lett.}}
  \textbf{\bibinfo{volume}{122}}, \bibinfo{pages}{010405}
  (\bibinfo{year}{2019}).
\newblock
  \urlprefix\url{https://link.aps.org/doi/10.1103/PhysRevLett.122.010405}.

\bibitem{periwal2021programmable}
\bibinfo{author}{Periwal, A.} \emph{et~al.}
\newblock \bibinfo{title}{Programmable interactions and emergent geometry in an
  array of atom clouds}.
\newblock \emph{\bibinfo{journal}{Nature}} \textbf{\bibinfo{volume}{600}},
  \bibinfo{pages}{630--635} (\bibinfo{year}{2021}).

\bibitem{sm2022}
\bibinfo{howpublished}{\url{URL_will_be_inserted_by_publisher}}.
\newblock \bibinfo{note}{For further details on the experimental setup, the
  theory formulation and simulation, and supporting experimental and
  theoretical results.}

\bibitem{futurePaper}
\bibinfo{author}{Guan, Q.} \emph{et~al.}
\newblock \bibinfo{title}{{Manuscript in preparation}}.

\bibitem{wu2000nonlinear}
\bibinfo{author}{Wu, B.} \& \bibinfo{author}{Niu, Q.}
\newblock \bibinfo{title}{Nonlinear {L}andau-{Z}ener tunneling}.
\newblock \emph{\bibinfo{journal}{Phys. Rev. A}} \textbf{\bibinfo{volume}{61}},
  \bibinfo{pages}{023402} (\bibinfo{year}{2000}).

\bibitem{guan2020nonexponential}
\bibinfo{author}{Q. Guan, M.K.H. Ome, T.M. Bersano, S. Mossman, P. Engels, and D. Blume},
\newblock \bibinfo{title}{Nonexponential tunneling due to mean-field-induced
  swallowtails}.
\newblock \emph{\bibinfo{journal}{Phys. Rev. Lett.}}
  \textbf{\bibinfo{volume}{125}}, \bibinfo{pages}{213401}
  (\bibinfo{year}{2020}).

\bibitem{Smerzi1997selftrapping}
\bibinfo{author}{Smerzi, A.}, \bibinfo{author}{Fantoni, S.},
  \bibinfo{author}{Giovanazzi, S.} \& \bibinfo{author}{Shenoy, S.~R.}
\newblock \bibinfo{title}{Quantum coherent atomic tunneling between two trapped
  {B}ose-{E}instein condensates}.
\newblock \emph{\bibinfo{journal}{Phys. Rev. Lett.}}
  \textbf{\bibinfo{volume}{79}}, \bibinfo{pages}{4950--4953}
  (\bibinfo{year}{1997}).
\newblock \urlprefix\url{https://link.aps.org/doi/10.1103/PhysRevLett.79.4950}.

\bibitem{raghavan1999josephson}
\bibinfo{author}{Raghavan, S.}, \bibinfo{author}{Smerzi, A.},
  \bibinfo{author}{Fantoni, S.} \& \bibinfo{author}{Shenoy, S.~R.}
\newblock \bibinfo{title}{Coherent oscillations between two weakly coupled
  {B}ose-{E}instein condensates: Josephson effects, $\ensuremath{\pi}$
  oscillations, and macroscopic quantum self-trapping}.
\newblock \emph{\bibinfo{journal}{Phys. Rev. A}} \textbf{\bibinfo{volume}{59}},
  \bibinfo{pages}{620--633} (\bibinfo{year}{1999}).
\newblock \urlprefix\url{https://link.aps.org/doi/10.1103/PhysRevA.59.620}.

\bibitem{qingze2021lmgdpt}
\bibinfo{author}{Guan, Q.} \& \bibinfo{author}{Lewis-Swan, R.~J.}
\newblock \bibinfo{title}{Identifying and harnessing dynamical phase
  transitions for quantum-enhanced sensing}.
\newblock \emph{\bibinfo{journal}{Phys. Rev. Research}}
  \textbf{\bibinfo{volume}{3}}, \bibinfo{pages}{033199} (\bibinfo{year}{2021}).
\newblock
  \urlprefix\url{https://link.aps.org/doi/10.1103/PhysRevResearch.3.033199}.

\bibitem{gerbier2019shapiro}
\bibinfo{author}{Evrard, B.}, \bibinfo{author}{Qu, A.},
  \bibinfo{author}{Jim\'enez-Garc\'{\i}a, K.}, \bibinfo{author}{Dalibard, J.}
  \& \bibinfo{author}{Gerbier, F.}
\newblock \bibinfo{title}{Relaxation and hysteresis near {S}hapiro resonances
  in a driven spinor condensate}.
\newblock \emph{\bibinfo{journal}{Phys. Rev. A}}
  \textbf{\bibinfo{volume}{100}}, \bibinfo{pages}{023604}
  (\bibinfo{year}{2019}).
\newblock \urlprefix\url{https://link.aps.org/doi/10.1103/PhysRevA.100.023604}.

\bibitem{Hoang2013stabilization}
\bibinfo{author}{T.M. Hoang, C.S. Gerving, B.J. Land, M. Anquez, C. D. Hamley, and M. S. Chapman},
\newblock \bibinfo{title}{Dynamic stabilization of a quantum many-body spin
  system}.
\newblock \emph{\bibinfo{journal}{Phys. Rev. Lett.}}
  \textbf{\bibinfo{volume}{111}}, \bibinfo{pages}{090403}
  (\bibinfo{year}{2013}).
\newblock
  \urlprefix\url{https://link.aps.org/doi/10.1103/PhysRevLett.111.090403}.

\bibitem{Hoang2016parametric}
\bibinfo{author}{Hoang, T.~M.} \emph{et~al.}
\newblock \bibinfo{title}{Parametric excitation and squeezing in a many-body
  spinor condensate}.
\newblock \emph{\bibinfo{journal}{Nat. Commun.}} \textbf{\bibinfo{volume}{7}},
  \bibinfo{pages}{11233} (\bibinfo{year}{2016}).
\newblock \urlprefix\url{https://doi.org/10.1038/ncomms11233}.

\bibitem{zhang2010localization}
\bibinfo{author}{Zhang, W.}, \bibinfo{author}{Sun, B.},
  \bibinfo{author}{Chapman, M.~S.} \& \bibinfo{author}{You, L.}
\newblock \bibinfo{title}{Localization of spin mixing dynamics in a spin-1
  {B}ose-{E}instein condensate}.
\newblock \emph{\bibinfo{journal}{Phys. Rev. A}} \textbf{\bibinfo{volume}{81}},
  \bibinfo{pages}{033602} (\bibinfo{year}{2010}).
\newblock \urlprefix\url{https://link.aps.org/doi/10.1103/PhysRevA.81.033602}.

\bibitem{fujimoto2019floquetspinor}
\bibinfo{author}{Fujimoto, K.} \& \bibinfo{author}{Uchino, S.}
\newblock \bibinfo{title}{Floquet spinor {B}ose gases}.
\newblock \emph{\bibinfo{journal}{Phys. Rev. Research}}
  \textbf{\bibinfo{volume}{1}}, \bibinfo{pages}{033132} (\bibinfo{year}{2019}).
\newblock
  \urlprefix\url{https://link.aps.org/doi/10.1103/PhysRevResearch.1.033132}.

\bibitem{PhysRevA.100.033617}
\bibinfo{author}{Li, Zheng-Chun and Jiang, Qi-Hui and Lan, Zhihao and Zhang, Weiping and Zhou, Lu},
  \newblock \bibinfo{title}{Nonlinear Floquet dynamics of spinor condensates in an optical cavity: Cavity-amplified parametric resonance},
\newblock \emph{\bibinfo{journal}{Phys. Rev. A}} \textbf{\bibinfo{volume}{100}},
  \bibinfo{pages}{033617} (\bibinfo{year}{2019}).
\newblock
  \urlprefix\url{https://link.aps.org/doi/10.1103/PhysRevA.100.033617}


\bibitem{feldmann2018qpt}
\bibinfo{author}{P. Feldmann, M. Gessner, M. Gabbrielli, C. Klempt, L. Santos, L. Pezze, and A. Smerzi},
\newblock \bibinfo{title}{Interferometric sensitivity and entanglement by
  scanning through quantum phase transitions in spinor {B}ose-{E}instein
  condensates}.
\newblock \emph{\bibinfo{journal}{Phys. Rev. A}} \textbf{\bibinfo{volume}{97}},
  \bibinfo{pages}{032339} (\bibinfo{year}{2018}).
\newblock \urlprefix\url{https://link.aps.org/doi/10.1103/PhysRevA.97.032339}.

\bibitem{mirkhalaf2020sensing}
\bibinfo{author}{Mirkhalaf, S.~S.}, \bibinfo{author}{Witkowska, E.} \&
  \bibinfo{author}{Lepori, L.}
\newblock \bibinfo{title}{Supersensitive quantum sensor based on criticality in
  an antiferromagnetic spinor condensate}.
\newblock \emph{\bibinfo{journal}{Phys. Rev. A}}
  \textbf{\bibinfo{volume}{101}}, \bibinfo{pages}{043609}
  (\bibinfo{year}{2020}).
\newblock \urlprefix\url{https://link.aps.org/doi/10.1103/PhysRevA.101.043609}.

\bibitem{rls2022optical}
\bibinfo{author}{Sundar, B.} \emph{et~al.}
\newblock \bibinfo{title}{Bosonic pair production and squeezing for optical
  phase measurements in long-lived dipoles coupled to a cavity}
  (\bibinfo{year}{2022}).
\newblock \eprint{arXiv:2204.13090}.

\end{thebibliography}
\end{document}


\title{Supplemental Material: Manipulation of nonequilibrium spin dynamics of an ultracold gas in a moving optical lattice
}

\author{Z.~N. Hardesty-Shaw}
\affiliation{Department of Physics, Oklahoma State University, Stillwater, Oklahoma 74078, USA}

\author{Q. Guan}
\affiliation{Homer L. Dodge Department of Physics and Astronomy, The University of Oklahoma, 440 W. Brooks Street, Norman,
Oklahoma 73019, USA}
\affiliation{Center for Quantum Research and Technology, The University of Oklahoma, 440 W. Brooks
Street, Norman, Oklahoma 73019, USA}
\affiliation{Department of Physics and Astronomy, Washington State University, Pullman, WA 99164, USA}

\author{J.~O. Austin}
\affiliation{Department of Physics, Oklahoma State University, Stillwater, Oklahoma 74078, USA}

\author{D. Blume}
\affiliation{Homer L. Dodge Department of Physics and Astronomy, The University of Oklahoma, 440 W. Brooks Street, Norman,
Oklahoma 73019, USA}
\affiliation{Center for Quantum Research and Technology, The University of Oklahoma, 440 W. Brooks
Street, Norman, Oklahoma 73019, USA}

\author{R.~J. Lewis-Swan}
\affiliation{Homer L. Dodge Department of Physics and Astronomy, The University of Oklahoma, 440 W. Brooks Street, Norman,
Oklahoma 73019, USA}
\affiliation{Center for Quantum Research and Technology, The University of Oklahoma, 440 W. Brooks
Street, Norman, Oklahoma 73019, USA}

\author{Y. Liu}
\affiliation{Department of Physics, Oklahoma State University, Stillwater, Oklahoma 74078, USA}

\date{\today}

\maketitle

\section{Experimental Details}

\subsection{Experimental sequence\label{sec:ExpSeq}}
Our experimental sequences start by creating a $S=1$ spinor BEC of up to $10^5$ sodium ($^{23}$Na) atoms in a crossed, anisotropic harmonic optical dipole trap (ODT) at a
particular quadratic Zeeman shift $q$ tuned by external magnetic fields, similar to our previous
work~\cite{chen2019quantum2,zhao2014dynamics,zhao2015antiferromagnetic}. We apply a resonant radio-frequency (RF) pulse to
prepare an initial state with fractional population $\rho_0 = \langle \hat{n}_0\rangle/N = 0.5$ in the $|S=1,m=0\rangle$
state and zero magnetization, $M = \langle \hat{n}_1 - \hat{n}_{-1} \rangle/N = 0$. We then adiabatically load the initial
state into a one-dimensional moving optical lattice. The lattice is constructed from two nearly orthogonal lattice beams
originating from a single-mode laser with wavelength $1064\,$nm and characterized by the
potential $V_{\rm lat}(\mathbf{r},t) = u_L \cos^2[\mathbf{k}_L \cdot \mathbf{r} - 2\pi\Delta f(t)t/4]$, with lattice
vector $\mathbf{k}_L$ oriented at approximately $40^{\circ}$  from the $z$-axis defined by gravity.  The resulting
standing wave potential has a lattice spacing of $\lambda_L/2=0.81\mu\mathrm{m}$. The time-dependent frequency difference
$\Delta f(t) = |f_H - f_V|$, where $f_H$ and $f_V$ are the corresponding lattice beam
frequencies, determines the velocity $v$ of the moving lattice, $v=\lambda_L (f_H - f_V)$. The velocity $v$ is manipulated
via a linear ramping rate $\alpha = \dfrac{h (\Delta f(t_2) - \Delta f(t_1))}{t_2 - t_1}$ such that when $v < 0$ ($v > 0$)
the atoms move in the $\mathbf{p}=2\hbar\mathbf{k}_L$ ($\mathbf{p}=-2\hbar\mathbf{k}_L$) direction. In the main text, the data in
Fig.~1 were taken with positive velocities, while the data in
Figs.~2 and 3 were taken
with negative velocities. The value of $\Delta f$ is initially set to zero and, after an adiabatic ramp of the lattice
depth $u_L$ to its final value at $t = t_1$, is quenched to its final value. The total time the atoms spend in the lattice is denoted by $t_F$ (for details see
Fig.~1(i) of the main text). At the
conclusion of each sequence, the trapping potentials are turned off so that the atoms can ballistically expand and be
captured using a two-step microwave imaging after a given time of flight (TOF)~\cite{chen2019quantum2}. Each data point in
this paper is an average of at least 8 repeated measurements and all error bars reported are estimated one standard
deviations.

\subsection{Optical dipole trap}
\label{sec_odt}

\begin{figure*}[tbh!]
	\includegraphics[width=100mm]{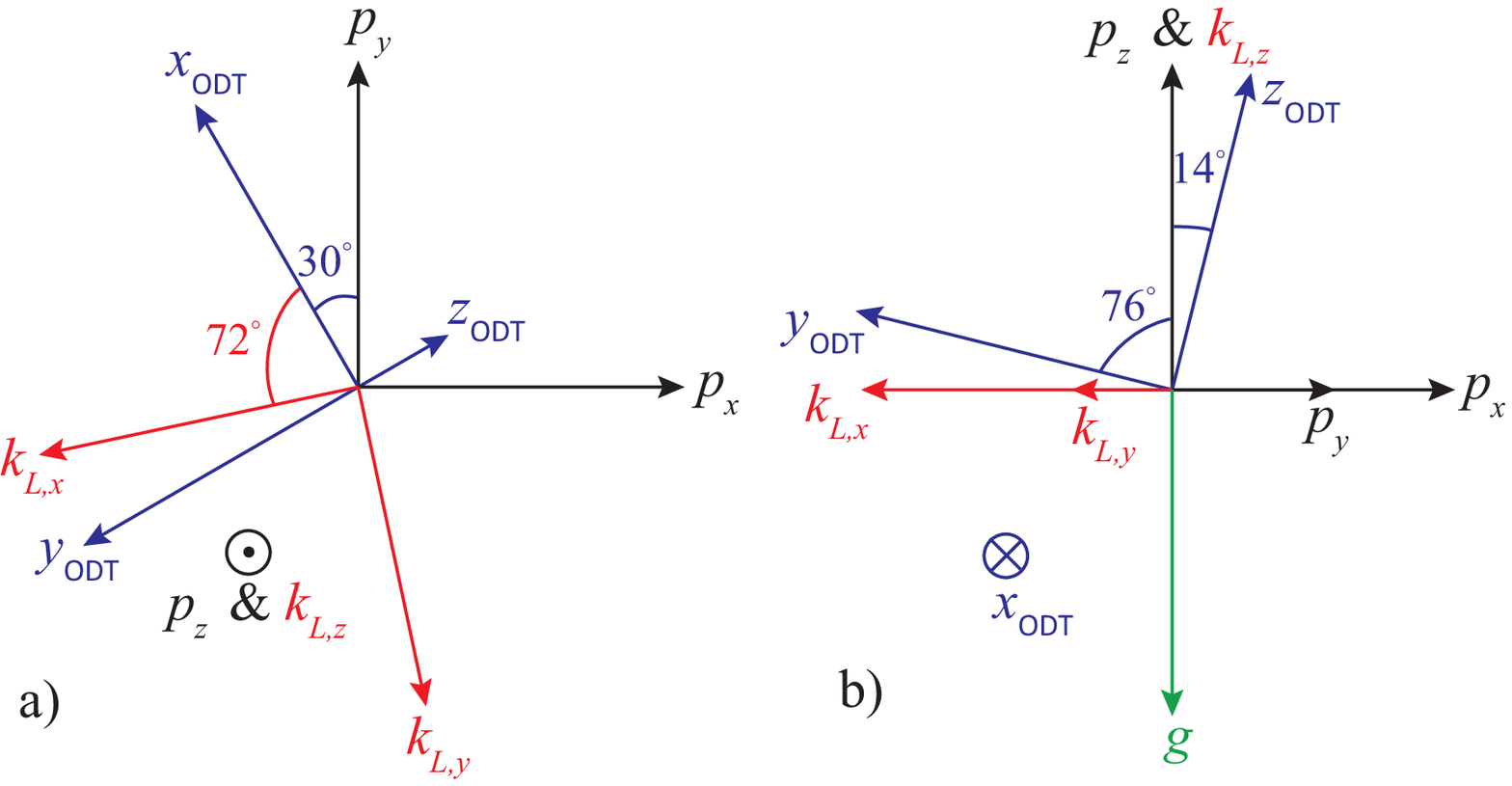}
	\centering
	\caption{\label{S-Geometry} a) The three coordinate systems used in our theoretical calculations projected into the 
	plane
	spanned by $p_x$ and $p_y$. Blue (red) [black] axes refer to the ODT (moving lattice, with associated lattice vector $\mathbf{k}_L$) [imaging plane, with associated momentum vector $\mathbf{p}$] coordinate system. b) Similar to a) but projected into the plane spanned by
	$y_{\rm ODT}$ and $z_{\rm ODT}$.  The green vector labeled by $g$ indicates the direction of gravity. The projections in this figure are to scale. }
\end{figure*}

An essential element of our experimental setup is a harmonic confinement potential skew to the moving lattice potential. The interplay of these two potentials triggers the nontrivial spatial dynamics that are key to our findings.  The harmonic confinement is provided by a crossed optical dipole trap (ODT) constructed by two orthogonal beams with wavelength $\lambda=1064~\mathrm{nm}$.  One ODT beam (ODT1) is orthogonal to gravity while the other (ODT2) is at a 76 degree angle relative to gravity (see Fig.~\ref{S-Geometry}). ODT1 is along the $x_{\rm ODT}$ axis, while the projection of ODT2 into the plane normal to gravity falls along the $y_{\rm ODT}$ axis (see Fig.~\ref{S-Geometry}).
The moving lattice lies 72 degrees horizontally from ODT1 and is tilted at a 40 degree angle relative to gravity.  Due to experimental considerations, our theoretical calculations therefore occur in three distinct coordinate systems that share a common $z$ axis defined by gravity: the coordinate systems defined by the ODT potential, the moving lattice, and the imaging plane, as illustrated in Fig.~\ref{S-Geometry}.

The potential generated by our crossed ODT can be parameterized to a good approximation by
\begin{eqnarray}
\label{eqn:ODT3D}
V_{\mathrm{3D}}(x_{\mathrm{ODT}},y_{\mathrm{ODT}},z)=-
V_0
\left\{
\frac{1}{1+\frac{x_{\mathrm{ODT}}^2}{z_0^2}}\exp\left[\frac{-2(y_{\mathrm{ODT}}^2+z^2)}{w_0^2\left(1+\frac{x_{\mathrm{ODT}}^2}{z_0^2}\right)}\right]+
\frac{1}{1+\frac{y_{\mathrm{ODT}}^2}{z_0^2}}\exp\left[\frac{-2(x_{\mathrm{ODT}}^2+z^2)}{w_0^2\left(1+\frac{y_{\mathrm{ODT}}^2}{z_0^2}\right)}\right]
\right \},
\end{eqnarray}
where $V_0=\frac{\mathcal{P}_0 h \alpha_{\rm fs}\lambda^2{\lambda_{\mathrm{Na}}}^2}{2\pi^3 m_e c^2 w_0^2  ( \lambda^2-{\lambda_{\mathrm{Na}}}^2)}$, $w_0=33~\mu\mathrm{m}$ is the ODT beam waist, $z_0=\frac{\pi w_0^2}{\lambda}$ is the associated Rayleigh length, $\mathcal{P}_0$ is the ODT power, $\lambda_{\mathrm{Na}}$ is the D2 line of sodium atoms, $\lambda$ is the wavelength of the ODT beam, $m_e$ is the mass of an electron, $h$ ($\hbar$) is the (reduced) Planck constant, $\alpha_{\rm fs}\approx\frac{1}{137}$ is the fine structure constant, $g$ is the gravitational acceleration, and $c$ is the speed of light. The ODT power, $\mathcal{P}_0$, can be varied to change the effective trap depth and size. 

\begin{figure*}[tbh!]
	\includegraphics[width=90mm]{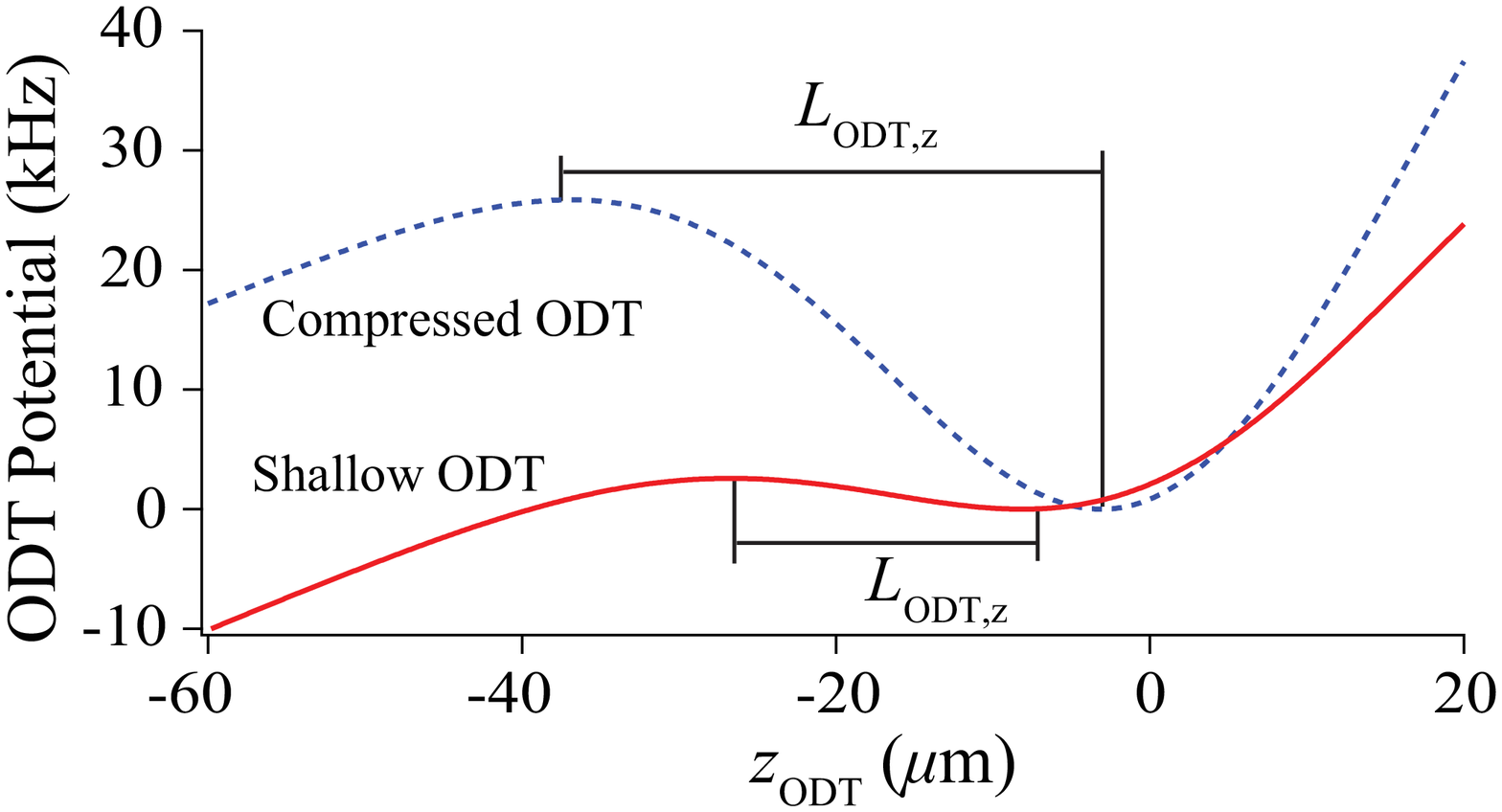}
	\centering
	\caption{\label{fig:figS2} 
		Cross-sectional cut of the ODT potential after accounting for gravity in the $z_{\rm ODT}$-direction at a local minimum in the $x_{\rm ODT}$ and $y_{\rm ODT}$ directions. The effective length of the ODT along the $z_{\rm ODT}$-direction is $L_{\rm ODT,z} = 20\mathrm{\mu m}$ for the shallow ODT (red solid line) and $L_{\rm ODT,z} = 33\mathrm{\mu m}$ for the compressed ODT (blue dashed line). Similar plots can be made for the ODT potential in the $y_{\rm ODT}$-direction, giving an effective length of $L_{\rm ODT,y} = 31\mathrm{\mu m}$ ($L_{\rm ODT,y} = 40\mathrm{\mu m}$) for the shallow (compressed) ODT.}
\end{figure*}

In Fig.~S2 we utilize Eq.~(\ref{eqn:ODT3D}) and take into account the effects of gravity to generate cross-sectional cuts for the compressed ODT trap ($\mathcal{P}_0\approx 35~\mathrm{mW}$), which was utilized for the experiments discussed in 
Fig.~3 of the main text and Fig.~\ref{fig:figSTOF} of the 
Supplementary Information,
and the shallow ODT trap ($\mathcal{P}_0\approx 17~\mathrm{mW}$), which was utilized 
for all other experimental figures and discussions. 
The effective trap length $L_{\rm ODT,i}$ is defined as the difference between the values for which $V_{3 \rm D}$ takes on a local maximum and local minimum (see Fig.~\ref{fig:figS2}) in the $i$-coordinate axis. Thus, the shallow trap (red solid line) is characterized by a 
smaller effective trap length $L_{\rm ODT,i}$ than the compressed trap (blue dashed line), i.e., atoms with high momentum exit the shallow trap more easily than the compressed trap. 
Since the trapping extends over a larger spatial region for the larger $L_{\rm ODT,i}$, the compressed trap extends the time scales over which  the complex spatial dynamics can be observed. However, the extended trapping times come at the cost of an increased average
atom temperature, which in turn tends to reduce the coherence and condensate fractions.
This is demonstrated in Fig.~\ref{fig:figSTOF}, which exhibits less coherent peaks than those shown in the analogous time-of-flight (TOF) image in Fig.~2(c) of the main text.

\begin{figure*}[tbh!]
	\includegraphics[width=60mm]{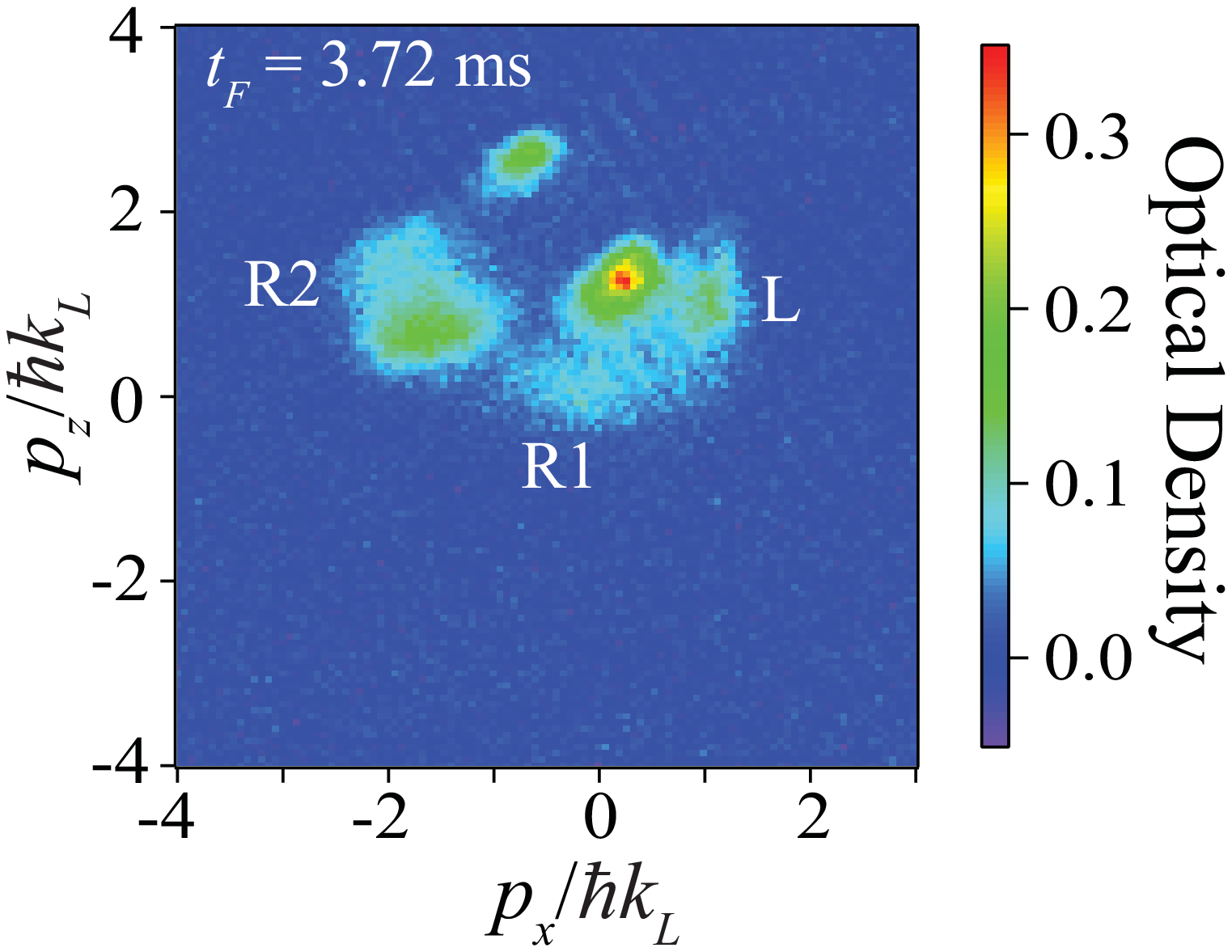}
	\centering
	\caption{\label{fig:figSTOF} Typical experimental TOF images obtained with the compressed ODT $(L_{\rm ODT,z} = 33\mathrm{\mu m},~L_{\rm ODT,y} = 40\mathrm{\mu m})$ displayed in Fig.~\ref{fig:figS2}.
		Data are taken following the same experimental sequence as in Fig.~2(c) of the main text, which instead used the shallow ODT $(L_{\rm ODT,z} = 20\mathrm{\mu m},~L_{\rm ODT,y} = 31\mathrm{\mu m})$. The colorbar scale on the right indicates the optical density of the image. 
	}
\end{figure*}

\subsection{Spin-mixing oscillations}

\begin{figure*}[htb]
	\includegraphics[width=170mm]{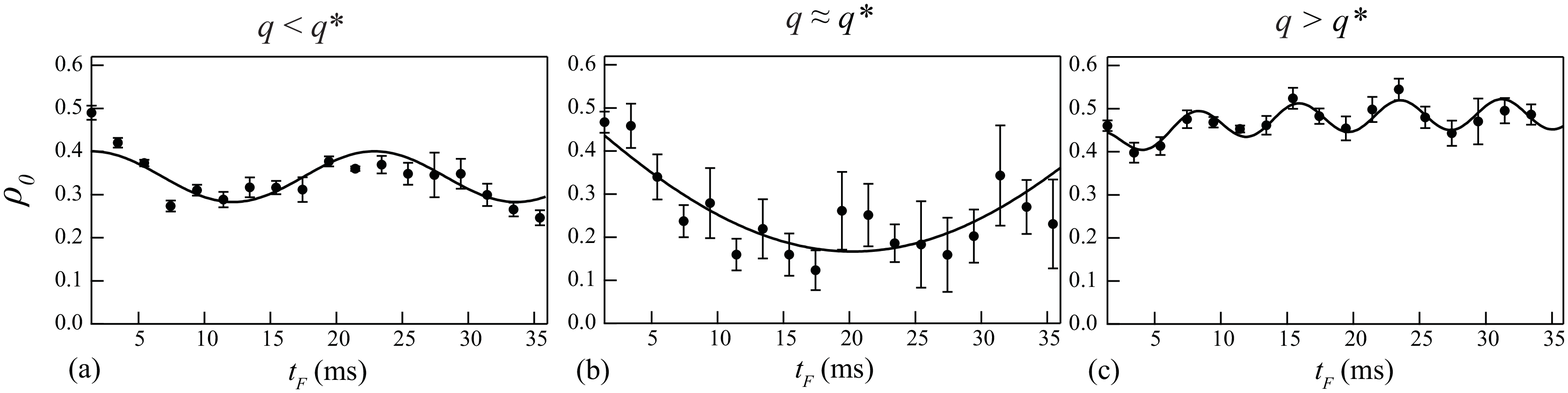}
	\centering
	\caption{\label{fig:PixelSum} Example time traces of $\rho_0$, the fractional population of the $|S=1,m=0\rangle$ state, obtained by counting all trapped atoms (see text) for: (a) $q/h = 15$~Hz, (b) $q/h = 25$~Hz, and (c) $q/h = 65$~Hz.  Here, $q$ is the quadratic Zeeman energy. All data is obtained from an experimental sequence illustrated in Fig.~1(i) of the main text with final lattice depth $u_L = 2.3E_R$ and fixed $\Delta f = 4.6E_R/h$. Sinusoidal fits (lines), which are used to extract the spin oscillation period T, are shown to guide the eye.
	}
\end{figure*}

All spin oscillation data presented in the main text are extracted from considering just the zero momentum, $\mathbf{p}=0$, peak. For completeness, Fig.~\ref{fig:PixelSum} 
presents spin oscillation data derived from the same data sets as Figs.~1(a)-(c) of the main text but obtained by counting all trapped atoms. Using this data, we obtain similar results to Figs.~1(a)-(c); specifically, the respective fitted periods agree within the margins of the errors.  The agreement between the distinct counting methods provides further support for the assumption of a dynamical single spatial-mode approximation (dSMA), as it indicates that coherence is retained between different momentum components as they evolve under the dynamics driven by the lattice potential.

\begin{figure*}[htb]
	\includegraphics[width=80mm]{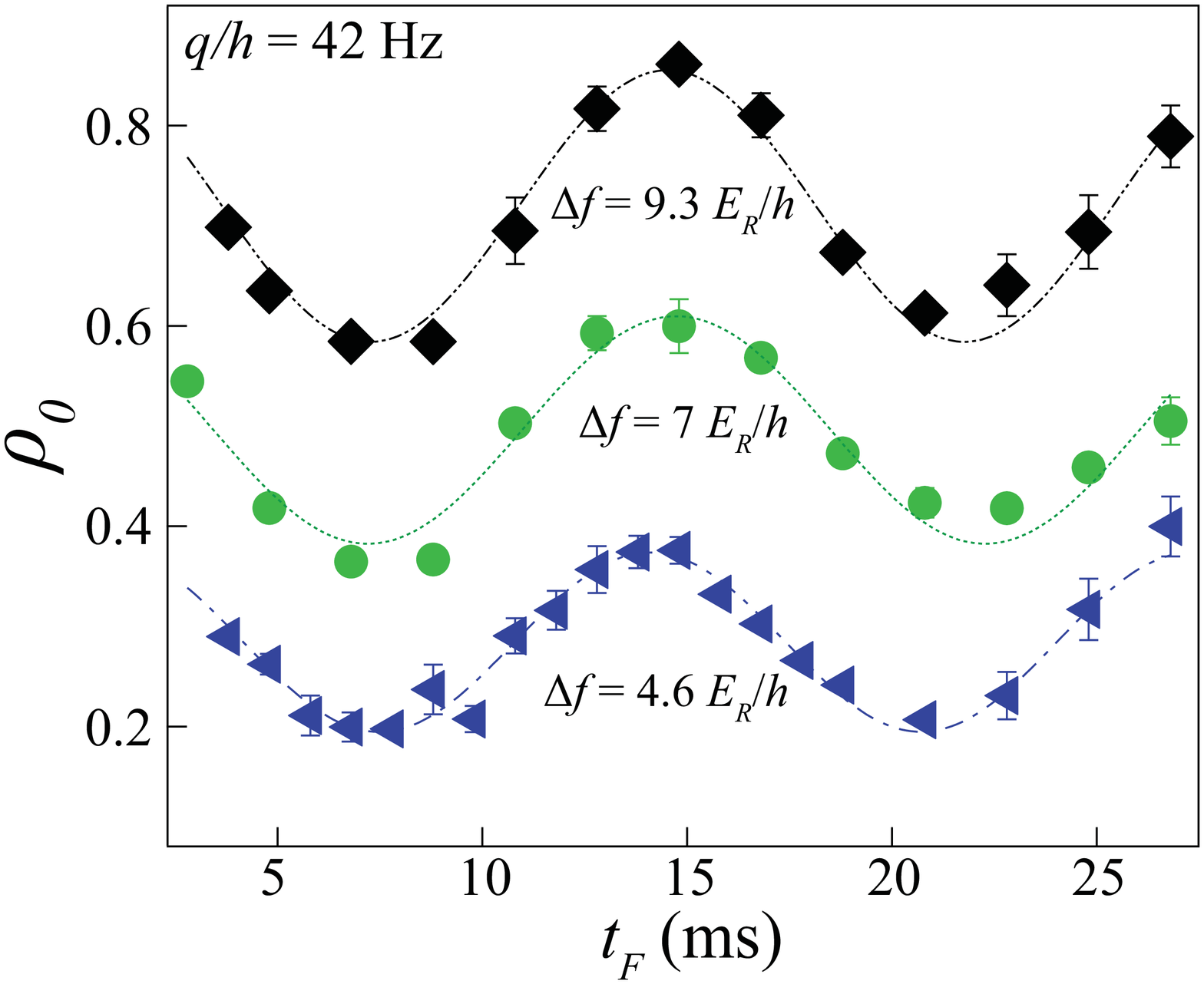}
	\centering
	\caption{\label{fig:figS4}  
	Time traces of $\rho_0$ at fixed $q/h = 42$Hz and a range of lattice speeds, set by $\Delta f = 4.6E_R/h$ (blue), $7E_R/h$ (green), and $9.3E_R/h$ (black). All data is for a final lattice depth $u_L = 2.3E_R$.
	Sinusoidal fits (lines), which are used to extract the spin oscillation period $T$, are shown to guide the eye. The $\Delta f = 7E_R/h$ ($\Delta f = 9.3E_R/h$) curves are offset by $0.2$ ($0.4$) in the $y$ direction for visual clarity.
	}
\end{figure*}

In Fig.~1(g) of the main text we see good agreement between the experimental data and the sSMA predictions in the Zeeman-dominated region. This is because we expect the dynamics of $c_2(t)$ to be less important in this region.
To support this, Fig.~\ref{fig:figS4} shows experimental data for time traces of $\rho_0$, the fractional population of the $|S=1,m=0\rangle$ state, obtained for a large range of $\Delta f$ at $q/h = 42$~Hz, in the Zeeman-dominated region. We observe no significant dependence of the period or amplitude of the oscillations on the actual value of $\Delta f$ (i.e., the speed of the moving lattice), and thus $c_2(t)$ associated with the resulting spatial dynamics.

\section{Details on theoretical treatment of spin-$1$ gases}

\subsection{Gross-Pitaevskii treatment of spin-$1$ condensates}
We consider a spin-1 condensate of $N$ sodium atoms of mass $M_{\text{Na}}$ under external confinement in the presence of a moving lattice. The two-body interactions are characterized by the spin-independent and spin-dependent interaction coefficients $g_0$ and $g_2$, 
\begin{eqnarray}
g_0=\frac{4\pi\hbar^2(a_{S=0} + 2a_{S=2})}{3 M_{\rm Na}}
\end{eqnarray}
and
\begin{eqnarray}
 g_2 = \frac{4\pi\hbar^2(a_{S=2} - a_{S=0})}{3 M_{\rm Na}},
 \end{eqnarray}
 where
$a_{S=0}=48.9 a_0$ and $a_{S=2}=54.5 a_0$ are, respectively, the $s$-wave scattering lengths for the $S=0$ and $S=2$ states with $a_0$ the Bohr radius \cite{Knoop2011PRA,chen2019quantum2}.
Our treatment includes the quadratic Zeeman shift term (proportional to $q$) but not the linear Zeeman shift, which does not play a role for the dynamics since it is conserved. 

At the mean-field level, the coupled dynamics of the spin and spatial degrees of freedom is described by the time-dependent spinor Gross-Pitaevskii (GP) equation,
\begin{widetext}
\begin{align}
\label{eqn:GP2D}
i\hbar\frac{\partial}{\partial t}\begin{pmatrix}
\psi_{-1}\\
\psi_0\\
\psi_1
\end{pmatrix}& =
\left[-\frac{\hbar^2\nabla^2}{2M_{\rm Na}} + V(\mathbf{r}, t) + g_0(N-1)\left(|\psi_{-1}|^2+|\psi_{0}|^2+|\psi_{1}|^2\right)\right]
\begin{pmatrix}
\psi_{-1}\\
\psi_0\\
\psi_1
\end{pmatrix}
+
\begin{pmatrix}
q & 0 & 0\\
0 & 0 & 0\\
0 & 0 & q
\end{pmatrix}
\begin{pmatrix}
\psi_{-1}\\
\psi_0\\
\psi_1
\end{pmatrix}\\\nonumber
& + g_2(N-1)\begin{pmatrix}
|\psi_{-1}|^2+|\psi_{0}|^2-|\psi_{1}|^2 & \psi_1^*\psi_0 & 0\\
\psi_1\psi_0^* & |\psi_{-}|^2+|\psi_{-1}|^2 & \psi_{-1}\psi_0^*\\
0 & \psi_{-1}^*\psi_0 & |\psi_{1}|^2+|\psi_{0}|^2-|\psi_{-1}|^2
\end{pmatrix}
\begin{pmatrix}
\psi_{-1}\\
\psi_0\\
\psi_1
\end{pmatrix} ,
\end{align}
\end{widetext}
where $V(\mathbf{r},t)$ includes contributions from the moving optical lattice $V_{\rm lat}(\mathbf{r},t)$, the confining potential (see Sec.~\ref{sec:ExpSeq}), and gravity. Within the GP formalism, each Zeeman component of the BEC is described by a mean-field wavefunction $\psi_m(\mathbf{r},t)$, such that both spin and spatial degrees of freedom, and their interplay, are simultaneously captured. The model ignores quantum fluctuations, which are expected to contribute minimally for the regimes in which the experiment operates.

For the scenarios reported in the main text, solution of the GP equation for the full $3$D system by direct numerical integration is not feasible. This is due to both the disparate timescales for the spin and spatial degrees of freedom as well as the large spatial region occupied by the BEC when it is kicked by the lattice. Specifically, the repeated momentum kicks imparted by the moving optical lattice on the fractured BEC lead to intricate spatial structure as well as a rapidly expanding cloud that must be tracked over comparatively long time scales (36~ms in Fig.~1 of the main text), requiring a large simulation box with good spatial resolution. Thus, the GP result presented in the main text are from numerical simulations of reduced dimensionality $1$D (Fig.~1) or $2$D (Fig.~2) models, wherein the interaction coefficients $g_0$ and $g_2$, mean spatial density of the BEC and trapping potential $V(\mathbf{r},t)$ are adjusted to capture the features of the experimental setup. The GP results presented are generated by numerically integrating the coupled GP equations using the XMDS2 software package \cite{DENNIS2013201}.

The dynamics shown in Figs.~2-3 of the main text are restricted to a few ms, i.e., much shorter times than those on which the spin oscillation dynamics occurs (see Fig.~1 of the main text). These dynamics out to 3~ms are amenable to 2D GP equations.
To this end, we  define a ``2D simulation plane"  by the direction of the $p_x-p_z$ imaging plane.  
The reduced dimensionality trap $V_{2 \rm D}$ is obtained by  
projecting the $3$D trap $V_{3 \rm D}$ onto the $2$D simulation plane.
Our simulations capture the role of the ODT trap (including, e.g., non-harmonic corrections to the confining potential) and its interplay with gravity.  Approximating the azimuthal and polar angles introduced in Sec.~\ref{sec_odt} by $30^\circ$ and $76^\circ$,
respectively, the $2$D trap is given by
\begin{eqnarray}
V_{\mathrm{2D}}(y,z)=V_{\mathrm{3D}}\left[\cos(76^\circ)\cos(30^\circ)y,-\sin(30^\circ)y,\sin(76^\circ)z\right].
\end{eqnarray}
The projection of the moving lattice in the $p_x-p_y$ plane is described by $k_{L,x}$ (see Fig.~\ref{S-Geometry}).
Neglecting the $12^{\circ}$ angle between the $k_{L,x}$ axis and the negative $p_x$ axis,
the moving lattice is in the $p_x-p_z$ plane.
Given the fact that the lattice has a $40^{\circ}$ angle relative to the $p_z$ axis (see the discussion above Fig.~\ref{S-Geometry}), the lattice vector is described by $\mathbf{k}_L=k_L[-\sin(40^{\circ}),0, \cos(40^{\circ})]$.
The reduction from $3$D to $2$D does change the mean density and, correspondingly, the chemical potential. This is accounted for by introducing the
effective $2$D interaction strength $g_{2\text{D}}^{\text{eff}}$;
similar to the 1D case,
$g_0(N-1)$ and $g_2(N-1)$
are replaced by $g_{2\text{D}}^{\text{eff}}$ and 
$g_{2\text{D}}^{\text{eff}}/28.06$, respectively.
The coupling constant $g_{2\text{D}}^{\text{eff}}$ is set by enforcing that the chemical potential of the $2$D system, obtained by solving the scalar GP equation, is equal to that of the $3$D system at $t=0$.

\begin{figure*}[ht]
\includegraphics[width=16cm]{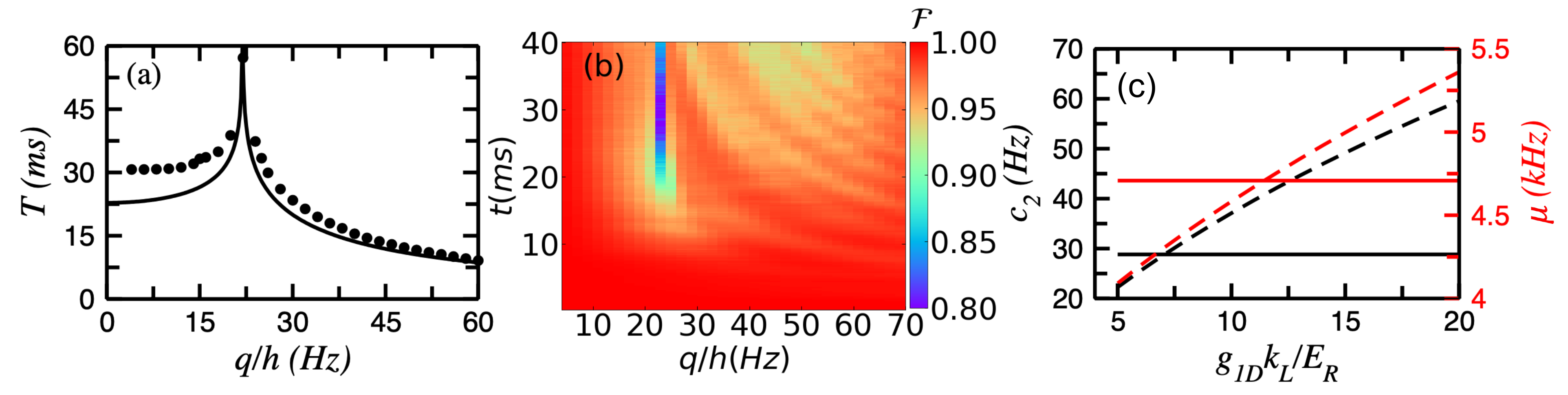}
\centering
\caption{\label{fig:figS6}
(a) Spin oscillation period $T$ as a function of Zeeman shift $q$. The black markers show $T$ extracted from $1$D spinor GP calculations that account for the spin-independent and spin-dependent interactions. The spin oscillation period diverges around $q^*/h = 22$~Hz. Emulating the analysis of the experimental data shown in Fig.~1(g) of the main text, the solid line is obtained by fitting the 1D GP data using the analytical sSMA expressions and treating the spin-spin interaction as a fitting parameter (namely, $c_{2,\mathrm{fit}}$). 
(b) Time evolution of the overlap of the Zeeman states (see Eq.~(\ref{eqn:ZeemanOverlap})) as a function of the Zeeman shift $q$. An overlap $\mathcal{F}$ of one indicates that all Zeeman states share a common spatial density profile, consistent with the dSMA. The only significant deviation from unity is observed for relatively long times ($t \gtrsim 20$~ms) when $q\approx q^*$.
(c) Parameter calibration of reduced dimensionality 1D GP equation. The black and red solid horizontal lines show the interaction energy $c_2$ (left axis) and chemical potential $\mu$ (right axis), respectively, for a 3D scalar BEC in the presence of a static lattice with $u_L=2.3$~$E_R$ for $N=80,000$ and angular trapping frequencies
$\omega_x=125$~Hz,
$\omega_y=125$~Hz,
and
$\omega_z=155$~Hz. The red and black dashed lines show $c_2$ and $\mu$, obtained by solving the 1D scalar
GP equation as a function of the 1D coupling constant $g_{1\text{D}}^{\text{eff}}$. 
The 1D calculations use $\omega_z=197$~Hz and $u_L=2.3$~$E_R$.}
\end{figure*}

The procedure for mapping the full 3D system to a reduced dimensionality model is not unique. Figure~\ref{fig:figS6}(c) illustrates this exemplary for the 1D case. Since the chemical potential $\mu$ is the characteristic energy scale of the time-independent  scalar 
GP equation and $c_2$ the characteristic energy scale of the effective spin Hamiltonian, it is natural to demand that the reduced dimensionality scalar GP model reproduces these two energy scales. The solid and dashed horizontal lines in Fig.~\ref{fig:figS6}(c) show the chemical potential $\mu$ and the spin-spin interaction strength $c_2$ of the 3D system in the presence of a static optical lattice with $u_L=2.3$~$E_R$ (note that the initial state preparation within the full 3D framework is significantly simpler than tracking the time evolution); this is the final lattice depth used in the experiment.
Fixing the parameters of the scalar 1D GP equation requires setting the values of $\hbar \omega_z/E_R$ and $g_{1\text{D}}^{\text{eff}}/(E_R k_L^{-1})$.
The 1D scalar GP equation is given by Eq.~(\ref{eq_gp2})
with the $\mathbf{r}$-vector replaced by $z$ and  $g_0(N-1)$ replaced by $g_{1\text{D}}^{\text{eff}}$, where $g_{1\text{D}}^{\text{eff}}$ has units of ``energy times length''.
The 1D spinor GP equation is obtained analogously, i.e., $g_0(N-1)$ and $g_2(N-1)$ are replaced by $g_{1\text{D}}^{\text{eff}}$ and $g_{1\text{D}}^{\text{eff}}/28.06$, respectively. 
The black and red dashed lines in Fig.~\ref{fig:figS6}(c) show $c_2$ and $\mu$
as a function of $g_{1\text{D}}^{\text{eff}}/(E_R k_L^{-1})$ for fixed $\hbar \omega_z/E_R$ 
(plugging in  $u_L=2.3$~$E_R$, this corresponds to $\omega_z=197$~Hz).
It can be seen that the dashed lines cross the solid lines at $g_{1\text{D}}^{\text{eff}}/(E_R k_L^{-1})\approx 7$
and $12.5$, respectively, i.e., for the $\hbar \omega_z/E_R$ value chosen there exists no unique $g_{1\text{D}}^{\text{eff}}/(E_R k_L^{-1})$ at which the values of $c_2$ and $\mu$ calculated within the 1D framework agree with the respective values calculated within the  3D framework. 
Since we do not find a unique $g_{1\text{D}}^{\text{eff}}/(E_R k_L^{-1})$  for other values  of $\hbar \omega_z/E_R$ either, there exists---unless additional conditions are added---an arbitrariness in the chosen 1D simulation parameters.
With this in mind, we select parameters that
place the divergence of $T$ at about the same $q$ value as observed experimentally (see Fig.~\ref{fig:figS6}(a)).
Specifically, the simulations shown in Figs.~1(d)-1(f) of the main text (and in Fig.~\ref{fig:figS6}(a)) use the same final lattice depth as the experiment (namely, $u_L=2.3$~$E_R$),
$\omega_z=197$~Hz, and $g_{1\text{D}}^{\text{eff}}/(E_R k_L^{-1})=12.0$.

\subsection{Single spatial-mode approximation and effective spin model}
We now motivate and introduce an approximate treatment that decouples the spin and spatial degrees of freedom.
Typically, the energy scales associated with the spin-independent terms of the Hamiltonian (i.e., the energy scales of the external harmonic and lattice confinement, the interactions that are proportional to $g_0$, and the chemical potential $\mu$ of the system) are of the order of kilohertz and much larger than those of the spin-dependent terms of the Hamiltonian (i.e., the value of the Zeeman shift $q$ and the interactions that are proportional to $g_2$), which are of the order of hertz.
This scale separation   motivates an approximate treatment wherein the spin and spatial degrees of freedom are treated independently~\cite{law1998spinmixing}, with the spatial degree of freedom controlled solely by the spin-independent terms of the Hamiltonian and the spin dynamics governed by the spin-dependent terms of the Hamiltonian.  

Following the literature~\cite{stamper2013spinor}, we make a single spatial-mode approximation (SMA) wherein the bosonic field operators are decomposed as 
\begin{equation}\label{eqn:psiSMA}
    \hat{\psi}_m(\mathbf{r}) = \hat{a}_m \phi(\mathbf{r},t) ,
\end{equation}
where $\hat{a}_m$ ($\hat{a}^{\dagger}_m$) is a bosonic operator that destroys (creates) a particle in Zeeman state $m$  in a spatial mode defined by the spatial mean-field wavefunction $\phi({\mathbf{r}},t)$, which is normalized to
one, i.e.,
$\int \text{d}^3 \mathbf{r} |\phi(\mathbf{r},t)|^2 =1$;
in Eq.~(\ref{eqn:psiSMA}), $m$ labels the Zeeman states ($m=0$ and $\pm 1$). The key assumption of the SMA is that $\phi({\mathbf{r}},t)$ is identical for all three Zeeman states.
The mean-field wave function $\phi({\mathbf{r}},t)$ is the solution to the 
 time-dependent scalar Gross-Pitaevskii (GP) equation 
\begin{align}
\label{eq_gp2}
i\hbar\frac{\partial \phi(\mathbf{r}, t)}{\partial t}
\approx
\left[-\frac{\hbar^2\nabla^2}{2M_{\mathrm{Na}}}+V(\mathbf{r}, t)+g_0(N-1)|\phi(\mathbf{r}, t)|^2\right]\phi(\mathbf{r}, t) ,
\end{align}
where $V({\mathbf{r}},t)$ is the same as Eq.~(\ref{eqn:GP2D}).
The spin dynamics, in turn, is governed by  the effective Hamiltonian
$\hat{H}_{\mathrm{eff}}(t)$,
\begin{equation}\label{eqn:Ham}
        \hat{H}_{\mathrm{eff}}(t) = \frac{c_2(t)}{2N}\hat{\mathbf{S}} \cdot \hat{\mathbf{S}} + q(\hat{a}^{\dagger}_1\hat{a}_1 + \hat{a}^{\dagger}_{-1}\hat{a}_{-1}) ,
\end{equation}
where the quantity $\hat{\mathbf{S}}$
denotes a collective spin operator. The time-dependent interaction strength $c_2(t)$,
\begin{equation}
\label{eq_c2_timedependent}
    c_2(t)= (N-1) g_2\int d^3\mathbf{r}|\phi(\mathbf{r}, t)|^4, 
\end{equation}
 is driven by the time dependence of the spatial mean-field wave function
 $\phi(\mathbf{r},t)$, i.e., the spin dynamics is governed---through the coefficient $c_2(t)$---by the spatial dynamics.
 In our experiment, the spatial dynamics is, to a large degree, induced by the moving optical lattice potential.
The  time dependence of the interaction coefficient  $c_2(t)$ is distinct from prior works (e.g., Refs.~\cite{yi2002single,zhao2014dynamics,liu2009quantum}), which 
assumed that the condensate is prepared in the ground-state of a static confining potential such that subsequent spatial motion is minimal and to a good approximation $\vert \phi(\mathbf{r},t) \vert^2 = \vert \phi(\mathbf{r},0) \vert^2$. 
For this reason, we use the distinguishing nomenclature of dynamical SMA (dSMA; time-dependent $c_2$) and static SMA (sSMA; time-independent $c_2$) for our and prior works, respectively. The latter is recovered from Eq.~(\ref{eqn:Ham}) by assuming $c_2(t) = c_2$.
To zeroth-order, the dynamics of the moving lattice experiments  during the first few milli-seconds is dominated by spatial dynamics. At later times,
however, the spin degrees of freedom become increasingly important as evidenced by the observation of spin oscillations (see Fig.~1 of the main text).

The dSMA is supported by our 1D spinor GP calculations. First, Fig.~1 of the main text shows good agreement between dSMA and GP predictions for $\rho_0(t)$. Second, we can explicitly validate the assumption that each Zeeman state occupies a single common spatial mode by computing the overlap, 
\begin{equation}\label{eqn:ZeemanOverlap}
    \mathcal{F} = \frac{\vert \int dx  \psi^*_0(x) \psi_1(x)\vert}{\sqrt{\int dx\vert \psi_0(x) \vert^2} \sqrt{\int dx\vert \psi_1 (x) \vert^2}} .
\end{equation}
The time evolution of this quantity over a range of Zeeman shifts is plotted in Fig.~\ref{fig:figS6}(b). We observe that $\mathcal{F}$ remains near unity across the interaction dominated regime ($q < q^*$) throughout the timescales we investigate (up to $40$~ms in the GP calculations, which is longer than the $30$~ms covered by the experiment). In the Zeeman regime, the overlap remains close to unity for $t \lesssim 20$~ms before minor deviations appear. As might be expected naively, a substantial breakdown of the single spatial-mode approximation occurs in a narrow region around the critical regime, $q\approx q^*$.

The experimental spin oscillation data in Fig.~1 of the main text are analyzed 
using the sSMA, i.e., the
mean-field equations associated with $\hat{H}_{\text{eff}}(t)$ for a time-independent spin-spin interaction coefficient. 
Specifically,
the spin oscillation period $T$ is extracted  by fitting the experimentally measured
fractional population $\rho_0(t)$ for various $q$ with sinusoidal functions, which provide good approximations to the spin oscillation dynamics away from the critical regime where the period diverges. All fits include at least one full period of oscillation. To determine the spin-spin interaction coefficient from the extracted periods, we perform a nonlinear least squares fit of the  $T$-versus-$q$ data using the analytical solutions to the mean-field equations associated with $\hat{H}_{\text{eff}}$~\cite{zhang2005coherent}. The fit uses $\rho_0(0)=0.5$ and 
$\theta(0)=0$ and treats $c_2$ as a free parameter (we denote the fit result by $c_{2,\text{fit}}$). While $\rho_0(0)$ is measured experimentally, $\theta(0)$ is not. However, based on the experimental sequence used to prepare the initial Zeeman populations we expect that $\theta(0)$ is equal to $0$. As reported in the main text, our fitting procedure yields $c_{2,\mathrm{fit}} = 23.7(1)$~Hz.

To gain additional insights, we perform an analogous analysis for the spin oscillation data obtained by solving the 1D spinor GP equations.
Specifically, we obtain $\rho_0(t)$ and the associated period $T$ by solving the spinor GP equations (see markers in Fig.~\ref{fig:figS6}(a)), and then extract $c_{2,\text{fit}}$
from the $T$-versus-$q$ data 
using the sSMA (solid line in Fig.~\ref{fig:figS6}(a)). For the 1D spinor GP simulations shown in Fig.~\ref{fig:figS6}(a), the spin oscillation period diverges at $q^*/h \approx 22$~Hz, i.e., at roughly the same
value of the Zeeman energy as in the experiment presented in the main text.
The qualitative agreement between the experimental and  theoretical analysis, particularly the consistency with the sSMA results, is encouraging and suggests that the 1D spinor GP simulations provide a qualitatively correct description of the moving lattice experiments. 

The GP simulations also enable us to better understand why the spin oscillation period extracted from the experimental data is consistent with the predictions of sSMA, even though the time traces show significant discrepancies. In the regime $q \leq q^*$, the sSMA theory predicts that the spin oscillation period should be strongly correlated with the spin-spin interaction strength \cite{sengstock2006spinresonance,zhang2005coherent,zhao2015antiferromagnetic}, which the GP results (see, e.g., Fig.~1(h) of the main text) predict to decay relatively slowly, compared to the typical timescales of the spin degree of freedom, apart from initial transient dynamics for $t\lesssim 3$~ms. This suggests that the spin oscillation period obtained from the experimental data for $q \leq q^*$ should be interpreted as being reflective of the characteristic scale of $c_2(t)$ over the experimental sequence. On the other hand, in the regime $q \gg q^*$ the oscillation period is expected to be dominated by the quadratic Zeeman shift. Thus, the experimentally observed spin oscillation periods are fit well by the sSMA predictions as the time dependence of $c_2(t)$ is less relevant for larger $q$.

\section{Classical interpretation of spatial dynamics: Lissajous curves}

\subsection{Model}
Figure~3 of the main text presents an analysis of the experimentally observed spatial dynamics of the condensate. Specifically, the plot tracks the positions of the L- and R1-peaks in momentum space. These peaks are extracted by fitting each momentum peak in a TOF image with a 2D Gaussian and extracting the fitted position. Positions are then converted to quasimomentum. The main text compared the experimental results to the predicted Lissajous curves. This section introduces the classical trajectory model that generates these Lissajous curves, using a single confined particle and initial conditions that are determined self-consistently from the properties of the experimental optical lattice and confining potential.

\begin{figure*}[ht]
 \includegraphics[width=85mm]{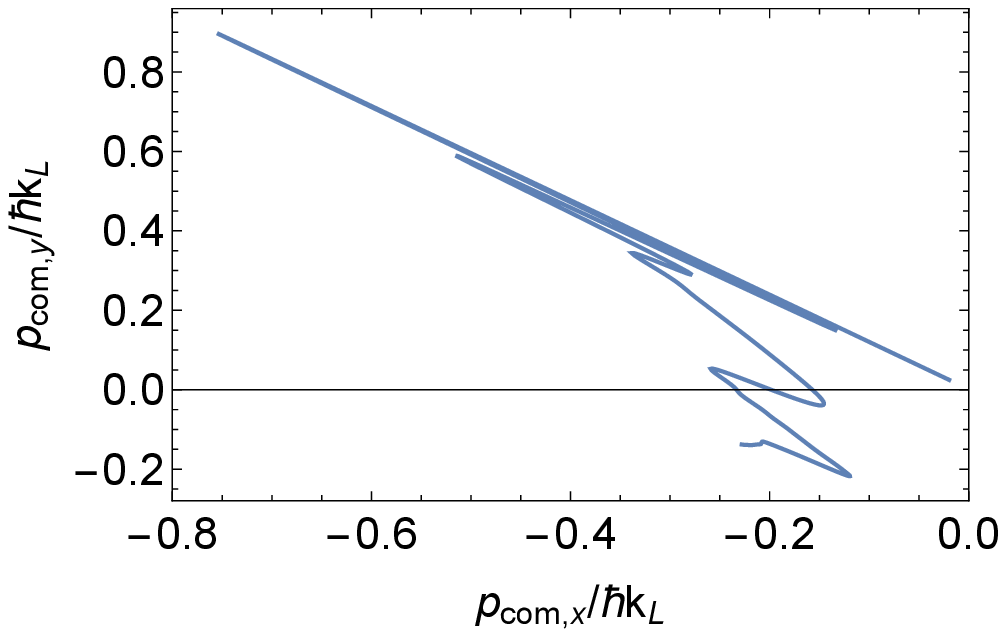}
 \centering
 \caption{\label{fig:COMMomentum} Example trajectory of the center-of-mass momentum $\mathbf{p}_{\rm com}$ extracted from a $2$D GPE simulation via $\mathbf{p}_{\rm com} = \int d^2\mathbf{p}~ \mathbf{p} \sum_m \vert \psi_m(\mathbf{p},t) \vert^2$,
 for a lattice depth of $u_L = 1.2E_R$ and $\Delta f = 4.6E_R/h$.
 }
\end{figure*}

Our model is underpinned by the insight that the spatial motion of the condensate atoms is predominantly controlled by the approximately harmonic trap and the main effect of the optical lattice is to stroboscopically impart sudden (instantaneous) kicks to the atoms when a resonance condition is satisfied dynamically.
This is well supported by an inspection of the time evolution of the center-of-mass (COM) momentum of the entire atomic cloud according to a $2$D GP calculation, as shown in Fig.~\ref{fig:COMMomentum}. The phase-space trajectory of the COM momentum is dominated by smooth evolution -- as different momentum components move freely within the confining potential $V_{\rm 2D}(x,z)$ -- with the exception of a sequence of regular abrupt reversals, indicating the sudden imparting of $2\hbar\mathbf{k}_L$ to a subset of atoms by the moving optical lattice. Our classical model constructs analogous Lissajous curves for an ensemble of individual (non-interacting) momentum components -- defined with respect to their initial position and velocity within the trap -- that are treated as classical particles traversing the $3$D confining potential $V_{\rm 3D}(\mathbf{r})$. Note that our classical trajectory simulations employ the full $3$D potential 
since these calculations are not numerically challenging. The initial position and momentum for each particle are determined by solving for the dynamically appearing resonances at which the moving lattice couples different momentum components. 

For concreteness, we introduce the specifics of our model by using it to describe the trajectory of the L-peak, which is assumed to be instantaneously populated at $t=t_1$ (i.e., when the lattice detuning is suddenly quenched) due to a near-resonant coupling mediated by the moving lattice between the initial BEC at $\mathbf{p}=0$ and a second non-zero momentum state $\mathbf{p} = 2\hbar \mathbf{k}_L$ whose kinetic energy differs by $4E_R/h \approx \Delta f$. Already, we note that the assumption that the L-peak is ``born'' instantaneously at $t=t_1$ is an oversimplification, but we argue that for lattice depths $u_L$ on the order of $E_R$ the coupling between momentum states generated by the moving lattice is much faster than any subsequent spatial motion.
Based on this description, the L-peak is initially located at the central minimum of the trap, which herein we define as the origin of our co-ordinate system, $\mathbf{r}_L(t=t_1)=(0,0,0)$, 
as it is directly outcoupled from the stationary BEC, and has initial momentum $\mathbf{p}_L(t=t_1)=2\hbar\mathbf{k}_L$. 

We expect the L-peak to quickly separate from the stationary BEC and then to begin climbing the walls of the confining potential and decelerate. To describe this evolution, we assume that the mean position and momentum of the L-peak follow a trajectory defined by an equivalent classical particle subject to the confining potential $V_{3\text{D}}$. Such a trajectory is described by solving the equations, 
\begin{equation}
 \label{eqn:classical}
     M_{\mathrm{Na}}\frac{d^2 \mathbf{r}(t)}{dt^2} = -\mathbf{\nabla} V_{\mathrm{3D}}\left[\cos(30^\circ)x-\sin(30^\circ)y,\cos(76^\circ)\sin(30^\circ)x+\cos(76^\circ)\cos(30^\circ)y+\sin(76^\circ)z,\sin(76^\circ)z-\cos(76^\circ)y\right]
\end{equation}
and
\begin{equation}
     \label{eqn:classical2}
     \mathbf{p}(t) =M_{\mathrm{Na}} \frac{d\mathbf{r}(t)}{dt}, 
\end{equation}
for the mean position and momentum of the corresponding classical particle, respectively. At this point, we highlight that we are assuming that the moving lattice is effectively invisible for the L-peak's motion; this is justified since the L-peak's motion rapidly changes its ``status'' from being on resonance with the  $\mathbf{p} = 0$ BEC at $t=t_1$ to being off-resonant with the  $\mathbf{p} = 0$ BEC for $t$ just a bit larger than $t_1$.

The L-peak moves freely within the confining potential until it dynamically satisfies a resonance condition where the optical lattice instantaneously couples the L-peak with momentum $\mathbf{p}_L(t)$ to a new state -- the R1-peak -- with momentum $\mathbf{p}_{R1}(t) = \mathbf{p}_{L}(t) + 2\hbar\mathbf{k}_L$. The resonance condition is defined by equating the difference between the kinetic energies of these states (it is assumed that the potential energy is unchanged as the new momentum component is created at the same location within the trap) with the energy imparted by the moving lattice, i.e., $4.6 E_R$, 
\begin{equation}\label{eqn:resonance}
    \frac{\vert \mathbf{p}_L(t) + 2\hbar\mathbf{k}_L \vert^2 }{2M_{\rm Na}} - \frac{\vert \mathbf{p}_L(t) \vert^2 }{2M_{\rm Na}}  = 4.6E_R.
\end{equation}
For the lattice vector $\mathbf{k}_L$ used in our moving-lattice experiment, this resonance condition can be written as a relation between different momentum components of the L-peak,
\begin{eqnarray}
 \label{eqn:resonance2}
 \mathbf{p}_{L}(t)\cdot\mathbf{k}_L=(0.6/4) \hbar k_L^2.
\end{eqnarray}
In the experiment, we approximately have $\mathbf{k}_L=k_L[-\sin(40^{\circ}),0, \cos(40^{\circ})]$, which leads to the resonance condition of $p_{L,z}=0.86p_{L,x}+0.20 \hbar k_L$.
The critical time when this resonance is fulfilled is estimated to be $t_{\rm cr} = 34.6\hbar/E_R = 1.67$~ms. The time $t_{\text{cr}}$ also defines the time at which we assume the R1-peak is ``born''. The initial condition for the R1-peak is then straightforward to compute as $\mathbf{r}_{R1}(t_{\mathrm{cr}}) = \mathbf{r}_{L}(t_{\mathrm{cr}}) = (-65.0, 0, 55.1)k_L^{-1}$ and $\mathbf{p}_{R1}(t_{\mathrm{cr}}) = \mathbf{p}_{L}(t_{\mathrm{cr}}) + 2\hbar\mathbf{k}_L = (-0.42, 0, -0.35)\hbar k_L + 2\hbar \mathbf{k}_L$. The computation of the Lissajous curve for the R1-peak, and subsequent creation of, e.g., the R2-peak, follows the same scheme as discussed above. 

We reiterate that these theoretical calculations should only be relied upon as a qualitative guide to help understand the mechanism of trap-induced resonances that lead to the experimentally observed dynamical fracturing of the BEC in momentum space. The Lissajous curves can only provide an approximate guide to the observed trajectories of the fractured components in momentum space for a number of reasons. Primarily, these are that the phase-space trajectories generated by Eqs.~(\ref{eqn:classical}) and (\ref{eqn:classical2}) ignore the role of: i) interactions and ii) time-dependent corrugation of the confining potential due to the moving lattice.

Lastly, we comment on the width of the approximate resonance. As previously discussed, the optical lattice couples momentum states separated by $2\hbar\mathbf{k}_L$ over a finite energy window, i.e., states that do not exactly fulfill the resonance criterion Eq.~(\ref{eqn:resonance2}). While quantifying the width of this energy window, or resonance region, is not straightforward, and further complicated by the role of interactions and motion of the BEC components in the ODT, we can provide a rough estimate. 

We use that the coupling of the initial $\mathbf{p} = 0$ BEC to the $\mathbf{p} = 2\hbar\mathbf{k}_L$ momentum state is off-resonance (the energy difference between these states is $4E_R$ rather than $4.6E_R$) to determine an estimate of a minimum width of the resonance region. 
Specifically, the fact that we see atoms transfer from the zero momentum state to  the $2 \hbar \mathbf{k}_L$ momentum state at $t \approx t_1$ for a detuning of $4.6E_R$, even though the true resonance for these momentum states occurs for a detuning of $4E_R$, indicates that the lattice-induced coupling of momentum states is near resonance for a window of at least $4.6E_R\pm 0.6E_R$. Correspondingly, the upper and lower boundaries of the resonance width can be derived by replacing the right hand side of Eq.~\eqref{eqn:resonance} with $5.2E_R$ and $4E_R$, respectively. This width is what is plotted in Fig.~3(b) of the main text and Fig.~\ref{fig:LissajousCurveCold} as gray shaded regions.

\subsection{\label{coldLissajous} Experimental comparison with Lissajous curves}

\begin{figure*}[ht]
 \includegraphics[width=75mm]{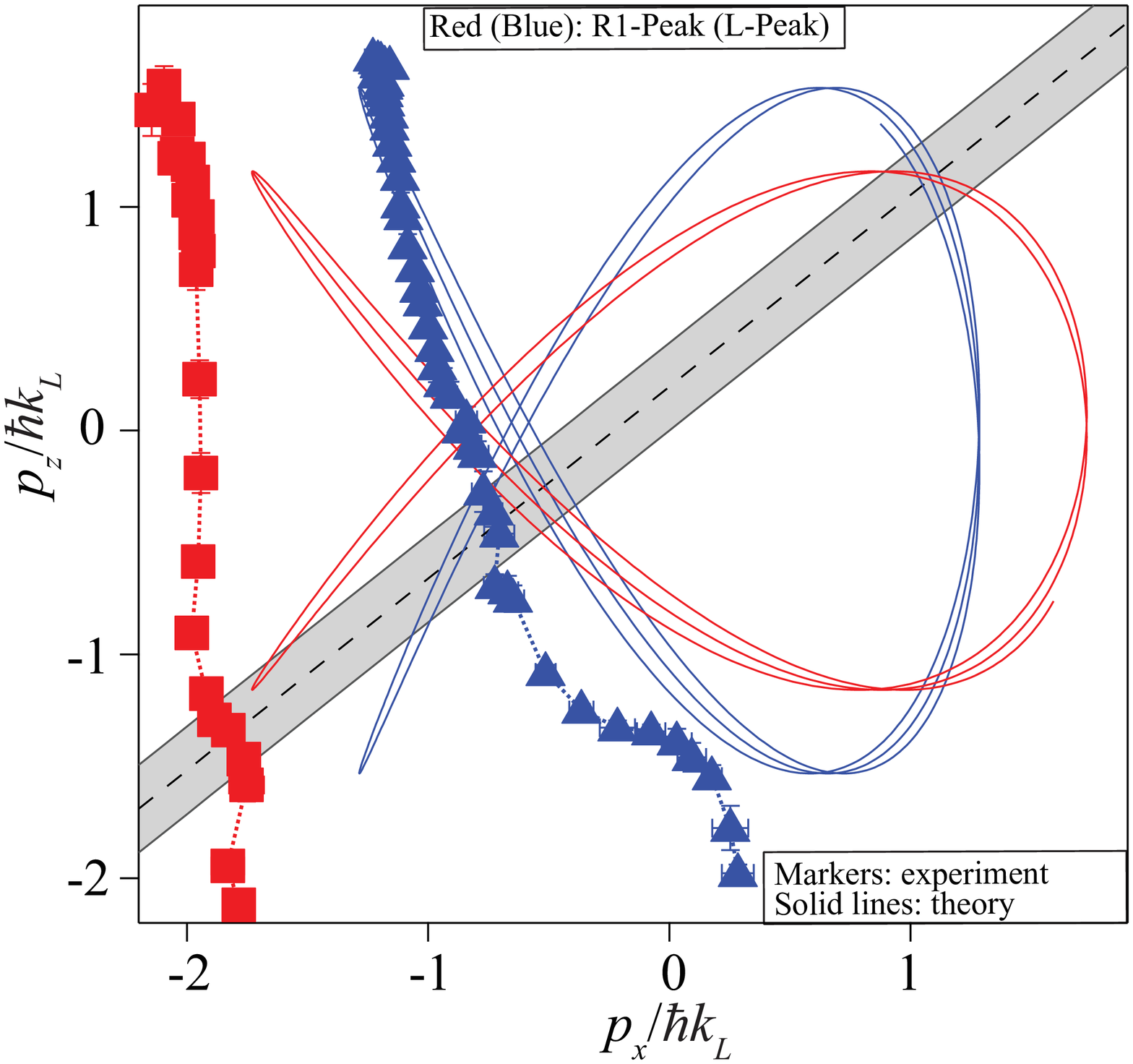}
 \centering
 \caption{\label{fig:LissajousCurveCold} Trajectories of the mean position of L-peak (blue triangles) and R1-peak (red squares), extracted similarly to Fig.~3 of the main text. Markers represent data taken in a typical ODT ($L_{\rm ODT,z} = 20~\mu$m).
Blue (red) solid lines are L-peak (R1-peak) positions according to Lissajous curves based on a calculation of a classical point particle in an ideal (effectively, $L_{\rm ODT,i} = \infty$) harmonic trap. Preceding L-peak formation, atoms occupy the $\mathbf{p}=0$ state at $p_z=p_x=0$ (prepared BEC) and the $\mathbf{p}=2\hbar \mathbf{k}_L$ state. The gray shaded region marks the approximate resonance region where decelerated atoms are kicked by the lattice (to create, e.g., the R1-peak).}
\end{figure*}

As illustrated in Fig.~3 of the main text, theoretical calculations of the Lissajous trajectories qualitatively reproduce the experimental trajectories that utilize the compressed ODT. However, as discussed in the main text, ODT compression is accompanied by lower condensate fractions and reduced visibility (see also Fig.~\ref{fig:figSTOF}). In Fig.~\ref{fig:LissajousCurveCold} we compare the Lissajous curves with the ODT typically employed in our experiment (uncompressed trap with $L_{\mathrm{ODT,z}} = 20~\mu$m). We find stark differences between experiment and theory curves for these parameters, including an early exiting of atoms out of the ODT trap due to gravity.
The disagreements found in the compressed trap of Fig.~3 and Fig.~\ref{fig:LissajousCurveCold} illustrate that fully capturing all the aspects of the experiment is necessarily beyond the simplified classical model, which is designed to minimally capture the dynamical appearance of resonances between different momentum states. A closer quantitative comparison would be provided by, e.g., a full 3D GP simulation and analysis of momentum space dynamics analogous to what is carried out in Figs.~3 and \ref{fig:LissajousCurveCold} for the experimental data.